\documentclass[12pt]{article}
\input epsf
\usepackage{hyperref}
\hypersetup{
    bookmarks=true,         
    unicode=false,          
    pdftoolbar=true,        
    pdfmenubar=true,        
    pdffitwindow=true,      
    pdfnewwindow=true,      
    colorlinks=true,       
    linkcolor=black,          
    citecolor=gray,        
    filecolor=magenta,      
    urlcolor=black          
}

\usepackage{amsmath}
\usepackage{amssymb}
\usepackage{stmaryrd}
\usepackage[dvips]{graphicx}
\usepackage{wrapfig}
\usepackage{cite}
\usepackage{verbatim}
\usepackage{setspace}
\usepackage{color}

 \usepackage[usenames,dvipsnames]{pstricks}
 \usepackage{epsfig}
 \usepackage{pst-grad} 
 \usepackage{pst-plot} 

 \usepackage{crop}

 \usepackage{bbm}

    \def\be{\begin{eqnarray}}
    \def\ee{\end{eqnarray}}
    \def\no{\nonumber}

    \def\suml{\sum\limits}
    \def\prodl{\prod\limits}

    \def\intii{\int\limits_{-\infty}^{\infty}}

    \def\bn{\begin{enumerate}}
    \def\en{\end{enumerate}}
    \def\bi{\begin{itemize}}
    \def\ei{\end{itemize}}

    \def\({\left(\!}
    \def\){\!\right)}
    \def\<{\left\langle\,}
    \def\>{\, \right\rangle}
    \def\[{\left[}
    \def\]{\right]}

    \def\bar{\overline}
    \def\hat{\widehat}

    \def\a{\alpha}
    
    \def\d{\delta}

    \def\l{\lambda}
    \def\s{\sigma}
    
    \def\th{\theta}
    
    \def\o{\omega}

    \def\CF{{\cal F}}
    \def\CG{{\cal G}}

    \def\CK{{\cal K}}
    \def\CL{{\cal L}}

    \def\CP{{\cal P}}
    \def\CQ{{\cal Q}}

    \def\pd{\partial}

    \def\MC{{\mathbb{C}}}

    \def\MI{{\mathbb{I}}}

    \def\MR{{\mathbb{R}}}

    \def\MZ{{\mathbb{Z}}}

    \def\sign{{\rm{sign}}\,}

    \def\su{\textit{su}}
    
    \def\psu{\textit{psu}}

   \def\gl{{gl}}
   \def\su{{su}}
   \def\sl{{sl}}
   \def\ualgebra{{u}}
   \def\groupn{\CN}
   \def\groupm{\CM}
   \def\groupk{\CK}
   \def\groupr{\CR}
    \def\glnm{\gl(\groupn|\groupm)}

   \def\gl{{\mathfrak{gl}}}
   \def\su{{\mathfrak{su}}}
   \def\sl{{\mathfrak{sl}}}
   \def\ualgebra{{\mathfrak{u}}}
   \def\groupn{\mathfrak{n}}
   \def\groupm{\mathfrak{m}}
   \def\groupk{\mathfrak{p}}
   \def\groupr{\mathfrak{r}}
    \def\glnm{\gl(\groupn|\groupm)}

    \def\cg{\mathfrak{g}}

    \def\inh{\chi}

 \def\xOOX{{\begin{picture}(63,10)
\thicklines
\put(45.8,-0.28){\bf\Large $\times$}
\put(5.15,4){\line(1,0){6.85}}
\put(17.5,4){\circle{11.3}}
\put(23.3,4){\line(1,0){6.75}}
\put(34.95,4){\circle{11.3}}
\put(41.45,4){\line(1,0){5.35}}
\put(52.5,4){\circle{11.3}}
\put(0,1.15){{$\mathbf{\times}$}}
\put(0,1.165){{$\mathbf{\times}$}}
\put(0,1.135){{$\mathbf{\times}$}}
\put(0.15,1.15){{$\mathbf{\times}$}}
\put(-0.15,1.15){{$\mathbf{\times}$}}
\end{picture}}}

 \def\xOXX{{\begin{picture}(63,10)
\thicklines
\put(45.8,-0.28){\bf\Large $\times$}
\put(28.8,-0.28){\bf\Large $\times$}
\put(5.15,4){\line(1,0){6.85}}
\put(17.5,4){\circle{11.3}}
\put(23.3,4){\line(1,0){6.75}}
\put(35.45,4){\circle{11.3}}
\put(41.45,4){\line(1,0){5.35}}
\put(52.5,4){\circle{11.3}}
\put(0,1.15){{$\mathbf{\times}$}}
\put(0,1.165){{$\mathbf{\times}$}}
\put(0,1.135){{$\mathbf{\times}$}}
\put(0.15,1.15){{$\mathbf{\times}$}}
\put(-0.15,1.15){{$\mathbf{\times}$}}
\end{picture}}}

\newcommand{\com}[1]{{}}

\begin{document}

\thispagestyle{empty}

\
\vspace{1.7cm}
\renewcommand{\thefootnote}{\fnsymbol{footnote}}
\setcounter{footnote}{0} \setcounter{figure}{0}
\begin{center}
{\Large\textbf{\mathversion{bold} String hypothesis for $\glnm$ spin chains: a particle/hole democracy.}\par}

\vspace{1.0cm}

\textrm{Dmytro Volin$^{}$}
\\ \vspace{1.2cm}
\footnotesize{

\vspace{3mm}

\textit{$^{}$ Department of Physics, The Pennsylvania State University\\ University Park, PA 16802, USA} \\
\texttt{dvolin AT psu.edu}
\vspace{3mm}

\textit{Bogolyubov
Institute for Theoretical Physics\\ 14b Metrolohichna Str.  \\
Kyiv, 03143 Ukraine} \\
\vspace{3mm}}

\par\vspace{1.5cm}

\textbf{Abstract}\vspace{2mm}
\end{center}
\noindent 

This paper is devoted to integrable $\glnm$ spin chains which allow for formulation of the string hypothesis. Considering the thermodynamic limit of such spin chains, we derive linear functional equations that symmetrically treat holes and particles. The functional equations  naturally organize different types of excitations  into a pattern equivalent to the one of Y-system, and, not surprisingly, the Y-system can be easily derived from the functional equations. The Y-system is known to contain most of the information about the symmetry of the model, therefore we map the symmetry knowledge directly to the description of string excitations. Our analysis is applicable for highest weight representations which for some choice of the Kac-Dynkin diagram  have only one nonzero Dynkin label. This generalizes known results for the AdS/CFT spectral problem and for the Hubbard model.
\small

\vspace*{\fill}

\setcounter{page}{1}
\renewcommand{\thefootnote}{\arabic{footnote}}
\setcounter{footnote}{0}

\newpage


\tableofcontents
\newpage


\section{Introduction}

\qquad Study of integrable systems at finite temperature or at finite volume often lead to the functional equations known as the Y-system \cite{Zamolodchikov:1991et}. This system had been constructed for most of the symmetry algebras with simply laced Dynkin diagrams, see for example  \cite{Gromov:2008gj} for a summarizing list. However, a supersymmetric generalization was not known until recently when it was proposed for the AdS/CFT integrable system \cite{Gromov:2009tv}. This proposal was then confirmed by explicit thermodynamic Bethe Ansatz (TBA)\cite{Zamolodchikov:1989cf} derivation in  \cite{Bomb,GromovKKV,Arutyunov:2009ur}.

\begin{figure}
\begin{centering}
\begin{tabular}{ccc}
\includegraphics[width=0.23\textwidth]{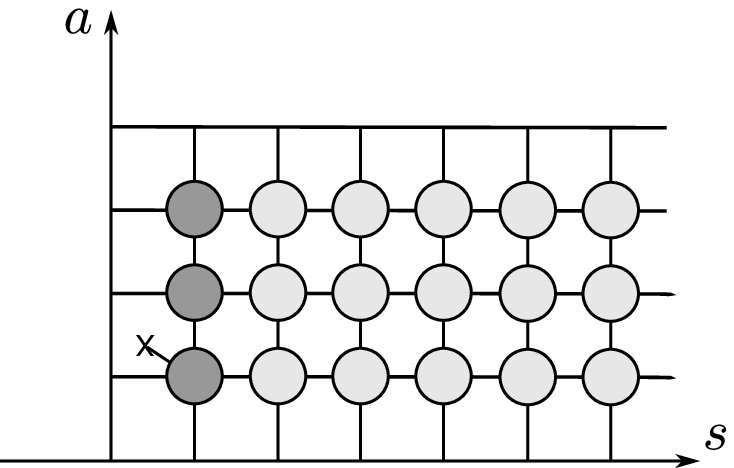} && \includegraphics[width=0.25\textwidth]{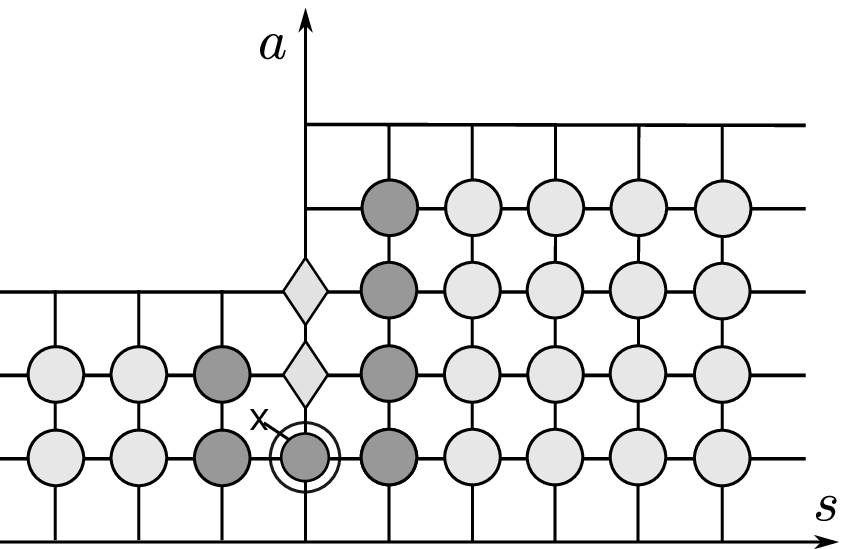} \\
a) $\su(\groupn)$, strip &\ \ \ \ & b) $\su(\groupn,\groupm)$, slim hook \\
\\
\includegraphics[width=0.2\textwidth]{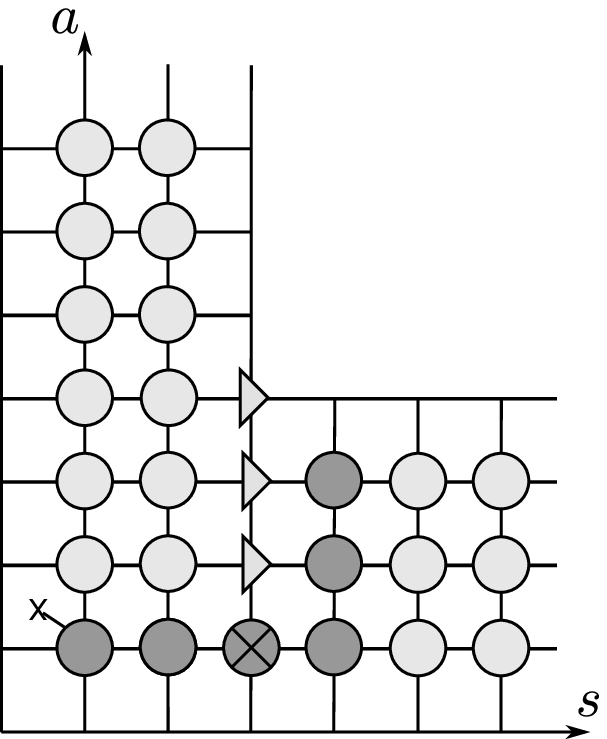}&&\includegraphics[width=0.3\textwidth]{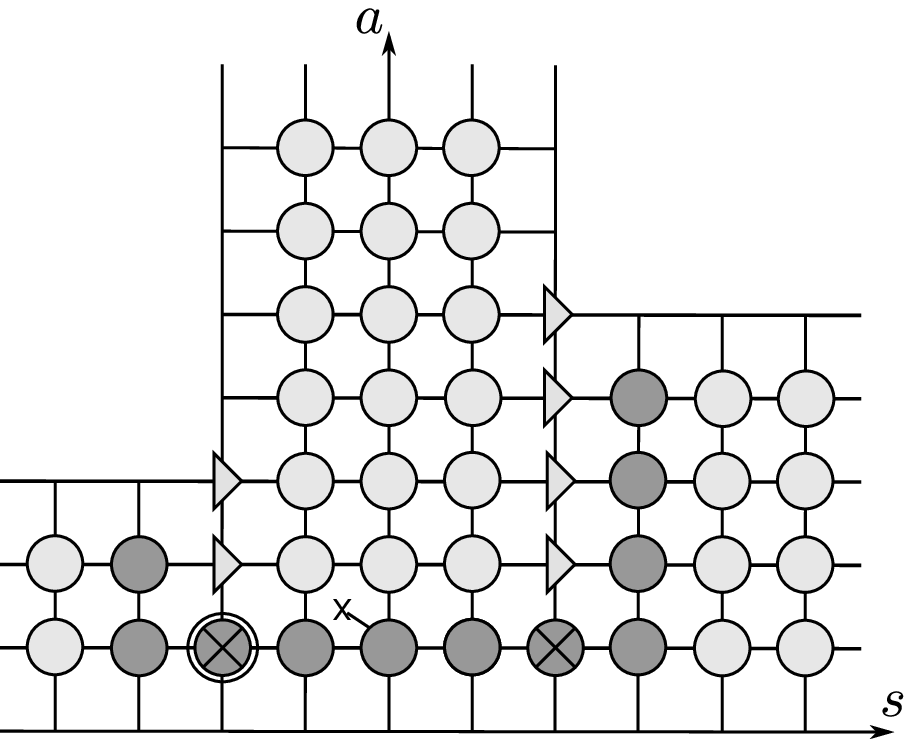}\\
c) $\su(\groupn|\groupm)$, fat hook && d)  $\su(\groupn|\groupm|\groupk)$, T-hook
\end{tabular}
\caption{\label{fig:patterns}Positions of different types of particles (string configurations) that define equation (\ref{mainequation}). The same domains appear for the Y-system (\ref{Ystandard}) and for the Hirota relations (\ref{Hirotaintro}) between the transfer matrices.}
\end{centering}
\end{figure}

Supersymmetric integrable systems were of interest well before the AdS/CFT case. The study of continuous limit of $\glnm$ rational spin chains first time was systematically approached in \cite{Saleur:1999cx}. Already in this paper a set of TBA equations was derived from which one can get in principle the Y-system on  a fat-hook (see Fig.~\ref{fig:patterns}). Such Y-system was however not obtained in \cite{Saleur:1999cx}.

In \cite{Saleur:1999cx} it was understood that, unlikely with a $\gl(\groupn)$ analog, $\glnm$ spin chains in covariant representations do not lead in the continuous limit to relativistic field theories.  One of the reasons for this is absence of the singlet state in  tensor product of covariant representations, hence the impossibility to introduce
antiferromagnetic vacuum. Therefore further study  in this domain was focused on  alternating spin chains \cite{Essler:2005ag,Candu:2010qj}. These spin chains are nonunitary, but as an advantage they do contain the singlet state.

The AdS/CFT integrable system revealed interest to the unitary supersymmetric spin chains. Though these spin chains do not posses an antiferromagnetic vacuum, the ferromagnetic description is the one that appropriate in the AdS/CFT case. The TBA construction leads us to the Y-system defined on a so called T-hook. This is a very simply looking system. But first, it should be supplemented with a set of analytical conditions \cite{Cavaglia:2010nm} which are not that simple, and we expect to simplify them. Second, the derivation of the Y-system is rather lengthy and not at all reflect the apparent algebraic structure of the final result. In fact, the Y-system is a kind of the statement about the symmetry of the model \cite{Gromov:2010vb}, thus there should be a more direct way for its derivation.

Mainly motivated by the search of the origins of the Y-system, in this paper we address the question - how general is the Y-system? For what kind of spin chains do we expect to get the structures similar to the one of AdS/CFT case? We state that the same kind of Y-systems can be derived for arbitrary $\glnm$ spin chains in a highest weight representation and arbitrary choice of the Kac-Dynkin diagram,  with condition that there is a choice of grading when only one Dynkin label is not zero\footnote{If this label is equal to one,  such representation is sometimes called fundamental. This terminology  will  not be used in this article  to avoid confusion. }.

The usual way to derive the Y-system is  to go through the whole TBA procedure. However we will show that a T-hook pattern can be spotted already at the first step of TBA, when one formulates the string hypothesis and writes down the set of linear integral equations to approximate the Bethe Ansatz. 

String hypothesis has a long story of application in the         Bethe Ansatz systems. It  already appeared in a seminal work of H.Bethe \cite{Bethe:1931hc} where it was successfully applied to support hypothesis about completeness of the Bethe Ansatz. Later on, it was
used to study spectrum of excitations of various spin chains around antiferromagnetic vacuum \cite{Faddeev:1981ip,Faddeev:1984ft}. The string hypothesis applied for AdS/CFT case in \cite{Arutyunov:2009zu} originated from the work of Takahashi  \cite{Takahashi:1972} on the Hubbard model, where stack configurations (triangles in this paper) were first time introduced. The string hypothesis was used in \cite{Saleur:1999cx} when continuous limit of $\glnm$ spin chain was considered.\\

The main claim of this paper is that the integral equations which are obtained in the thermodynamic limit and under assumption of string hypothesis can be rewritten as in a remarkably symmetric form:
\be\label{mainequation}
        \sum_{a'}K_{a,a'}*\rho_{a',s}+\sum_{s'}K_{s,s'}*\rho_{a,s'}^*={\rm source\ term}.
\ee
These equations cover all the cases of integrable chains with a symmetry group from the A series. Up to our knowledge, similar symmetric formulation, in a Fourier space, first time appeared in \cite{PZJ} for $\gl(\groupn)$ case, here we make its supersymmetric generalization.

The operator  $K_{a,a'}$ is decomposed to
\be
        K_{a,a'}=\delta_{a,a'}K-\delta_{a,a'+1}-\delta_{a,a'-1},
\ee
where definition of $K$ requires some explanation and is discussed shortly below.

$\rho_{a,s}$ and $\rho_{a,s}^*$ denote the density functions for particles and their holes\footnote{In the horizontal strips of Fig.~\ref{fig:patterns} $\rho$ is assigned to particles, and $\rho^*$ - to holes, while the assignment in the vertical strip  is reversed. In the central region assignment rule depends on a particular choice of a Kac-Dynkin diagram.}, and we label different types of particles by two integer indices $\{a,s\}$. Exact rule for assigning $\{a,s\}$ labels to different particle types is given in the main text. This rule leads to organization of excitation types into one of the patterns
of Fig.~\ref{fig:patterns}, depending on a real form of  a symmetry algebra.

Equation (\ref{mainequation}) is satisfied everywhere inside of one of the domains of Fig.~\ref{fig:patterns}. There is an additional equation in the corners of fat- and T-hooks which is roughly obtained by inversion of $K_{a,a'}$ and $K_{s,s'}$ and is given by (\ref{cornerequation}).

Equation (\ref{mainequation}) already appeared in the PhD thesis of the author \cite{Volin:2010cq}, however with an indigestible proof and without complete understanding of the structure of equations in the corner. Also the relation between different choices of Kac-Dynkin diagrams was not clarified. In this paper with the help of the fermionic duality transformations we resolve all the complications that appeared in \cite{Volin:2010cq}.

To define function $(K*\rho)(u)$ we first introduce  the resolvent $R$ of $\rho$:
\be\label{resoldef}
        R(u)=\intii dv\frac {\rho(v)}{u-v},\ \ \  -\frac 1{2\pi i}(R(u+i0)-R(u-i0))=\rho(u), \ee
and functions
\be
        \CK{_\pm}R=(D+D^{-1})R,\ \ \ {\rm Im}(u)\gtrless 1/2,
\ee
where $D=e^{i/2\pd_u}$ is a shift operator.
The functions $\CK_+R$ and $\CK_-R$ are defined in the domains ${\rm Im}(u)>1/2$ and ${\rm Im}(u)<-1/2$ respectively and analytical there; then they can be continued to the whole complex plane,  the result may  depend on the chosen path.  $K*\rho$ is defined by a difference of $\CK_+R$ and $\CK_-R$:

\be
 K*\rho=-\frac 1{2\pi i}\left(\CK_+R-\CK_-R \right).
\ee

If $\rho(u)$ is an analytic function in the strip $|{\rm Im}(u)|\leq 1/2$ than we get
\be
        (K*\rho)(u)=\rho\left(u+\frac i2\right)+\rho\left(u-\frac i2\right)
\ee
for real values of $u$. However, if $\rho(u)$ has branch points in this strip than analytical continuation of $\CK_{\pm}R$ is not unique and $K*\rho$ would be a multivalued function.

As was discussed in \cite{Volin:2010cq}, from equations (\ref{mainequation}) the $Y$-system can be straightforwardly written:
\be\label{YK}
        \left(1+\frac 1{Y_{a',s}}\right)^{K_{a,a'}*}=\left(1+{Y_{a,s'}}\right)^{K_{s,s'}*},
\ee
where $Y_{a,s}=\frac{\rho_{a,s}^*}{\rho_{a,s}}$ is evaluated for the configuration that minimizes the free energy.

If $Y_{a,s}$ is analytic in the strip $|{\rm Im}(u)|\leq 1/2$, which is so for relativistic models,\footnote{Though in this paper we consider spin chains, relativistic models for which $Y$-system is written may be obtained as a continuous limit of  spin chains. Thus the equations (\ref{mainequation}) are the same for relativistic models after we exchange $\rho$ and $\rho^*$ for momentum-carrying excitations.} than (\ref{YK}) reduces to a standard formulation of the Y system\footnote{notation $f^\pm(u)=f(u\pm i/2)$ is used.}$^,$\footnote{We were informed that independently V.~Kazakov, A.~Kozak, and P.~Vieira obtained Y-system (\ref{Ystandard}) for a general T-hook pattern. Due to relation between (\ref{YK}) and (\ref{mainequation}) this is the same observation we discuss in this article. For $\su(2,2|4)$ case this is of course the observation of \cite{Gromov:2009tv}.}:
\be\label{Ystandard}
        Y_{a,s}^+Y_{a,s}^-=\frac{(1+Y_{a,s+1})(1+Y_{a,s-1})}{(1+Y_{a+1,s}^{-1})(1+Y_{a-1,s}^{-1})}.
\ee
For the AdS/CFT case some of $Y$-functions have cuts in the mentioned strip or on its border, therefore equations (\ref{Ystandard}) are satisfied only for the particular choice\footnote{for some small $a$ and $s$ additional kernels that depend on the 't Hooft coupling constant should be added to $K_{a,a'}$ and $K_{s,s'}$. However, for the particular choice of analytical continuation this does not affect (\ref{Ystandard}).} of analytical continuation of functions $\CK_\pm R$. A possibility to choose different prescriptions for continuation is a source of seeming discrepancy between  works \cite{Arutyunov:2009ur,Arutyunov:2009ux} and \cite{Gromov:2009tv,GromovKKV}. Let us note that all possible pathes should be chosen in order to define completely the Y-system while (\ref{Ystandard}) corresponds only to one of the pathes. Other choices should lead to monodromy constraints on the $Y$-functions formulated in \cite{Cavaglia:2010nm}.

\

Domains of Fig.~\ref{fig:patterns}, obtained here through the analytical study of sting hypothesis, have a quite different algebraic interpretation. One can assign so called rectangular representations (see sections \ref{sec:rectangular} and \ref{sec:unitary}) to each node of these domains in a way that transfer matrices in these representations would satisfy Hirota bilinear equations:
\be\label{Hirotaintro}
        T_{a,s}^+T_{a,s}^-=T_{a,s+1}T_{a,s-1}+T_{a+1,s}T_{a-1,s}.
\ee
The construction is well defined in the case of spin chains. In the case of relativistic models it was built explicitly only for some particular examples \cite{Bazhanov:1996aq}, and this allowed to find spectrum of the excited states\footnote{Ideology of TBA \cite{Zamolodchikov:1989cf} allows for computation of ground state only, though there are analytical continuation tricks \cite{Dorey:1996re} which allow computing energies of some excited states. Another approach is to derive nonlinear integral equations starting from the lattice regularization \cite{Destri:1992qk,Destri:1994bv} which can be generalized to compute energies of excited states \cite{Fioravanti:1996rz}. }.

Besides rectangular representations, there is a way to define the generalization of the Young tableaux that can be inscribed into a T-hook such that it defines a unitary highest weight representation (UHW) of the symmetry algebra \cite{Gromov:2010vb}. In this article we argue that all UHW can be obtained in this way.

\

The article is organized as follows. In section \ref{sec:notations} we introduce notations for $\glnm$ algebra, define the Bethe Ansatz equations and recall realization of fermionic duality transformations on the Bethe Ansatz equations. Section \ref{sec:mainresult} contains the derivation of our main result - equation (\ref{mainequation}). In section \ref{sec:unitary} we explain how to parameterize UHW in terms of generalized Young tableaux and conjecture that this allows to describe all UHW. We discuss obtained results and outline possible directions for future study in section \ref{sec:conclusions}. Article contains also three appendices, first two are devoted to showing some evidence supporting the string hypothesis. The third one shows how to treat more involved situations with the source term.

\section{\label{sec:notations}Integrable supersymmetric spin chains}
This section collects known in the literature results about Bethe Ansatz for rational spin chains. Since in this paper we are dealing with the string hypothesis, we will need basic knowledge about representations which allow for this hypothesis. These are so called rectangular representations, name coming from the shape of partition that encodes them.

\subsection{Conventions for $\glnm$ algebra.}

The $\glnm$ algebra is described using its generators $E_{ij}$, $1\leq i,j\leq \groupn+\groupm$, which obey the supercommutation relations:
\be
       [E_{ij},E_{kl}]=\d_{jk}E_{il}-(-1)^{(|i|+|j|)(|k|+|l|)}\d_{li}E_{kj}.
\ee
It is useful to think about $E_{ij}$ as the basis vector in the space $V\otimes V^*$, where $V=\MC^{\groupn+\groupm}$ is a graded vector space.  Than $|i|$ means the grading of the basis vectors $v_i$ in $V$ and can be 0 (even) or 1 (odd).

There are different ways to distribute gradings in $V$. We are not going to make some distinguished choice. Instead, we use the Kac-Dynkin diagram to define the grading.  If the $i$-th node of the Kac-Dynkin diagram is white then $v_i$ and $v_{i+1}$ have the same grading, if this node is crossed - then $v_i$ and $v_{i+1}$ have the opposite grading. Together with agreement what is the grading of $v_1$, this uniquely defines $|i|$. Then, the generators $E_{ij}$ are odd if $|i|\neq |j|$ and even otherwise.

Once the basis vectors $v_i$ are chosen, we define Borel decomposition of the $\glnm$ algebra be requiring that $E_{i,j}$ with $i<j$ are raising operators (correspond to positive roots). This will be called a standard Borel decomposition with respect to a corresponding grading.

The irreducible highest weight representation is uniquely defined by
\be
        E_{ij}|\Omega\rangle&=&0,\ \ i<j,\no\\
        E_{ii}|\Omega\rangle&=&m_i|\Omega\rangle.
\ee
To denote the weight of the irrep we will use the square brackets: $[m_1,\ldots,m_{N+M}]$.

To render formulas simple,  the hatted notations will be also used:
\be
        \hat E_{ii}=(-1)^{|i|}E_{ii},\ \ \hat m_i=(-1)^{|i|}m_i.
\ee
The general rule is that we will put hats when factors of type $(-1)^{|i|}$ are used in the definition and no hats otherwise.

The Cartan generators $h_i$ of the $\sl(\groupn,\groupm)$ subalgebra and correspondingly the Dynkin labels $\o_i$ are defined by\footnote{Some of the Dynkin labels may differ by sign from those defined in \cite{Frappat:1996pb}.}
\be
\begin{array}{rcccl}
\hat h_i &\!\!\! =\!\!\! & (-1)^{|i|}h_i &\!\!\!=\!\!\!& \hat E_{ii}-\hat E_{i+1,i+1}\,, \\
\hat \o_i&\!\!\! =\!\!\! & (-1)^{|i|}\o_i&\!\!\!=\!\!\!& \hat m_i-\hat m_{i+1}\,. \\
\end{array}
\ee
Set of the Dynkin labels will be enclosed in the angle brackets: $\langle \hat \o_1,\ldots,\hat \o_{N+M-1} \rangle$.

Note that negative value of $\hat\o_i$ would still mean compactness if $|i|=|i+1|=1$. And inverse, if $|i|=|i+1|=1$ and $\hat\o_i>0$, the corresponding representation is infinite dimensional.

\subsection{Bethe Ansatz equations}
Solution of the integrable spin chain by the inverse scattering method is based on the procedure of  diagonalization of the transfer matrix. The latter can be diagonalized by means of the (nested) algebraic Bethe Ansatz, once each site of the spin chain is in the highest weight representation.

The eigenvectors are parameterized by the solutions of the Bethe Ansatz equations. These equations for a generic $\glnm$ spin chain are given by \cite{Kulish:1983rd,Ragoucy:2007kg}:
\be\label{Bethegeneral}
        \prod_{k=1}^{\groupn+\groupm-1}\prod_{j=1}^{M_k}\frac{u_{i}^{(\ell)}-u_{j}^{(k)}+\frac i2{\<\a_\ell,\a_k\>}}{u_{i}^{(\ell)}-u_{j}^{(k)}-\frac i2{\<\a_\ell,\a_k\>}}=-\frac{\Lambda_\ell(u_i^{(\ell)}-\frac i2c_\ell)}{\Lambda_{\ell+1}(u_i^{(\ell)}-\frac i2c_\ell)}\ .
\ee
Here $\ell=\overline{1,\groupn+\groupm-1}$, $i=\overline{1,M_\ell}$ and the following notation is introduced
\bi

\item  $\Lambda_\ell$ is defined by
\be
        \Lambda_\ell(u)=e^{i\hat\phi_\ell}\prod_{k=1}^L (u-\inh_k+i\,\hat m_k^{(\ell)}).
\ee
\item $L$ is the length of spin chain.
\item A grading of the $\glnm$ and its Borel subalgebra are chosen. $\a_k$ are the simple positive roots with respect to this choice. The scalar product between simple roots (graded Cartan matrix) is given by:
\be
        \<\a_\ell,\a_k\>=\delta_{\ell,k}((-1)^{|\ell|}+(-1)^{|\ell+1|})-(-1)^{|\ell+1|}\delta_{\ell+1,k}-(-1)^{|\ell|}\delta_{\ell-1,k}.
\ee
\item $M_k$ is the number of Bethe roots at the Dynkin node $k$.
\item $[\hat m_k^{(1)},\ldots,\hat m_{k}^{(\groupn+\groupm)}]$ is the weight of the highest weight vector at spin chain cite $k$.
\item $c_\ell$ are defined by
\be
c_{\ell}=\sum_{k=1}^{\ell}(-1)^{|k|}\,.
\ee
\item $\inh_k$ are the inhomogeneity parameters - arbitrary numbers which are assigned to each node of the spin chain.

\item $e^{i\hat \phi_{\ell}}$ are the twist parameters. They appear if we insert in the supertrace which defines the transfer matrix the term $\CG={\rm diag}(e^{i\phi_1},\ldots,e^{i\phi_{\groupn+\groupm}})$, where $e^{i\hat\phi_{\ell}}=(-1)^{|\ell|}e^{i\phi_{\ell}}$.

Twists and inhomogeneities
are auxiliary parameters. Variation of them allows keeping solution of the Bethe Ansatz equations in a general position. We will use dependence on twist for study of string hypothesis in appendices \ref{sec:applicability} and \ref{sec:numerical}. \com{Dependence on twist is useful for a proof of completeness of the Bethe Ansatz \cite{jlksadf}.}
\ei

The weight $[\mu_1,\ldots,\mu _{\groupn+\groupm}]$ of the eigenvector associated with the solution of the Bethe Ansatz equations is given by
\be
        \mu_i=M_{i-1}-M_{i}+\,\sum_{k=1}^L m_k^{(i)}.
\ee
The eigenvector is the highest weight state in the case when $\phi_\ell=0$. Adding Bethe roots at infinity corresponds to the  action by the lowering operator. When the twists are turned on, the Bethe roots at infinity become finite\footnote{When the twists are present, notion of highest weight is not applicable since the symmetry algebra is reduced to the Cartan subalgebra only.}.

\subsection{\label{sec:duality}Duality transformations}

Consider change of the basis in $V$:
\be
        \sigma_k: v_i\mapsto v_{\sigma_k(i)},
\ee
where $\sigma_k(k)=k+1,\ \sigma_k(k+1)=k$ and $\sigma_k(i)=i$ for $i\neq k,k-1$.

This transformation changes the standard Borel subalgebra and therefore changes notion of the highest weight representation. So, operator $e_+=E_{k,k+1}$ transform to $e_-=E_{k+1,k}$. The highest weight vector $|\Omega\rangle$ remains such only in the case if
\be
        h_k |\Omega\rangle=0\,.
\ee
In all other cases $|\Omega\rangle$ does not remain the highest weight, however representation itself may remain of the highest weight type. Let us consider different possibilities

\paragraph{Compact case.} If $|k|=|k+1|$ and $\o_k$ is a nonnegative integer then $\su(2)$ subalgebra, generated by $e_\pm$ and $h_k$, forms a finite dimensional representation by acting on $|\Omega\rangle$. It is straightforward to see that the  vector $e_-^{\omega_k}|\Omega\rangle$ becomes a highest weight vector with respect to the standard Borel subalgebra of $\s_k(\glnm)$. The weight of $e_-^{\omega_k}|\Omega\rangle$ in correspondence with $\s_k(h)$ is the same as the weight of $|\Omega\rangle$ in correspondence with $h$ (here $h$ is the Cartan subalgebra).

\begin{figure}[t]
\centering
\includegraphics[width=10cm]{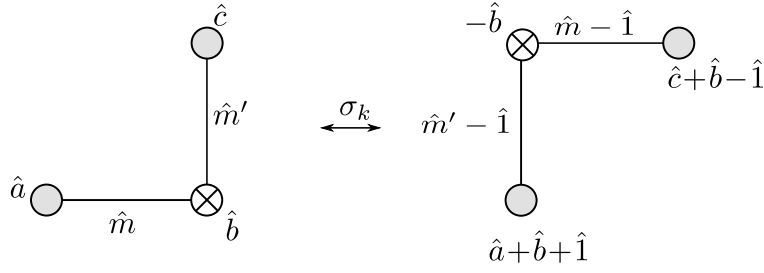}
\caption{\label{fig:duality} Fermionic duality transformation for the case $b\neq 0$. In the case $b=0$ $a,b,c$ do not change. Horizontal direction corresponds to $E_{ii}$ generators with $|i|=0$, vertical - with $|i|=1$.}
\end{figure}

\paragraph{Fermionic case.} If $e_\pm$ are fermionic generators ($|k|\neq |k+1|$) then together with $h_k$ they form an $\su(1|1)$ subalgebra. By acting on $|\Omega\rangle$ it forms two-dimensional representation. And it is straightforward to check that $e_-|\Omega\rangle$ is the highest weight vector with respect to the standard Borel decomposition of $\s_k(\glnm)$. Transformation of the highest weight, assuming that $|k|=0$ and $|k+1|=1$, is
given by\be\label{dualityweightchanging}
     [\ldots,\hat m_{k-1},\hat{m}_k,\hat{m}_{k+1},\hat m_{k+2},\ldots]\to [\ldots,\hat m_{k-1},\hat{m}_{k+1}-\hat1,\hat{m}_{k}-\hat 1,\hat m_{k+2},\ldots].
\ee
On the level of the  Kac-Dynkin diagram the transformation is shown in Fig.~\ref{fig:duality}

\paragraph{Noncompact case.} If $|k|=|k+1|$ and $\o_k<0$ or non-integer, the highest weight representation of the $su(2)$ subalgebra is infinite dimensional and becomes of lowest weight type after $\s_k$ transformation. For the whole $\glnm$ algebra, representation,  which is of highest weight type for the standard Borel decomposition with respect to initial grading, is no more of highest or lowest weight type after transformation $\s_k$ applied to a noncompact bosonic node.

\

Let us now consider how the automorphism $\sigma_k$ is realized on the level of the Bethe Ansatz equations.

\paragraph{Fermionic duality.}
First consider the case when $e_\pm$ are fermionic. The transformation known as a fermionic duality transformation was studied number of times in the literature \cite{Woynarovich,tJmodel,Tsuboi:1998ne,GohmannSeel}.

Since we heavily rely on the fermionic duality transformation in this text, we will repeat basic facts about it. Assume the grading as is in the l.h.s. of Fig.~\ref{fig:duality}, labeling of Bethe roots as in (\ref{uthvlabeling}) and consider Bethe equations for $u,\theta,v$ nodes. To be definite we take $u,v$ to be bosonic nodes, though this is not of importance. The Bethe equations are written as
\begin{subequations}
\label{betheforduality}
\be
       \label{first22} \frac{Q_{u_{-2}}^-Q_{u}^{++}Q_{\theta}^-}{Q_{u_{-2}}^+Q_{u}^{--}Q_{\theta}^+}&=&-\frac{\Lambda_{u_{-2}}^+}{\Lambda_u^+}\,,\\        \label{second22} \frac{Q_u^-}{Q_u^+}\frac{Q_v^+}{Q_v^-}&=&-\frac{ \Lambda_{u}}{\Lambda_\theta}\,,\\
       \label{third22} \frac{Q_\theta^+Q_v^{--}Q_{v_2}^+}{Q_\th^-Q_v^{++}Q_{v_2}^{-}}&=&-\frac{\Lambda_{\theta}^+}{\Lambda_{v_2}^+}\,.
\ee
\end{subequations}
Here $Q_\a$ are the Baxter polynomials, the first equation is evaluated at zeroes of $Q_u$, the second -- of $Q_\theta$, and the third -- of $Q_v$.

It is possible to simultaneously translate all the Bethe vectors by a constant. This allows changing all $c_\ell$ by a constant. We use this freedom to put $c_\theta=0$. Therefore $c_u=c_v=-1$. This explains the shifts by $i/2$ in $\Lambda$ in first and third equations of (\ref{betheforduality}).

Equation (\ref{second22}) can be rewritten as a $QQ$ relation:
\be\label{QQwithsource}
 \Lambda_\theta Q_u^-Q_v^++\Lambda_uQ_u^+Q_v^-=Q_\theta Q_{\overline\theta}\,(e^{i\hat\phi_\theta}+e^{i\hat\phi_{u}})\,.
\ee
The requirement that $Q_{\overline\theta}$ is a polynomial is equivalent to the Bethe Ansatz equations (\ref{second22}). Note now that $Q_\theta$ and $Q_{\overline\theta}$ enter in a symmetric way in (\ref{QQwithsource}), thus a natural idea is to exchange the role of them.

First, one can write the Bethe equations for the zeroes of $Q_{\overline\theta}$. These equations are exactly the same as (\ref{second22}). To preserve the pattern of (\ref{Bethegeneral}) we write it as
\renewcommand{\theequation}{\arabic{equation}b}
\be\label{Betheafterduality1}
 \frac{Q_u^+}{Q_u^-}\frac{Q_v^-}{Q_v^+}&=&-\frac{\Lambda_{\theta}}{\Lambda_u}\,.
\ee
This equation is evaluated at zeroes of $Q_{\overline\theta}$ - dual fermionic roots.

Let us replace, using (\ref{QQwithsource}), $Q_\theta$ with $Q_{\overline\theta}$ everywhere in (\ref{first22}) and (\ref{third22}) . For each application of (\ref{QQwithsource}) when replacing, only one of the terms in the l.h.s. of (\ref{QQwithsource}) survives and the result is the following:

\addtocounter{equation}{-1}
\renewcommand{\theequation}{\arabic{equation}}
\begin{subequations}
\label{Betheafterdualall}
\be
        \frac{Q_{u_{-2}}^-Q_{\overline\theta}^+}{Q_{u_{-2}}^+Q_{\overline\theta}^-}&=&-\frac{\Lambda^+_{u_{-2}}}{\Lambda_\theta^-}\,, \ee
\end{subequations}
\vspace{-1.3em}
\addtocounter{equation}{-1}
\renewcommand{\theequation}{\arabic{equation}c}
\be        \label{Betheafterduality2}  \frac{Q_{\overline\theta}^-Q_{v_2}^+}{Q_{\overline\th}^+Q_{v_2}^{-}}&=&-\frac{\Lambda_{u}^-}{\Lambda_{v_2}^+}\,.
\ee
Equations (\ref{Betheafterdualall}) precisely match (\ref{Bethegeneral}) for the choice of weights given by the r.h.s. of (\ref{dualityweightchanging}).

\renewcommand{\theequation}{\arabic{equation}}
If $\Lambda_\theta=\Lambda_u$, the $\Lambda$-s cancel out from  (\ref{second22}) and therefore r.h.s. of (\ref{betheforduality}) remains unaffected after duality transformations. This corresponds to the case $h_k|\Omega\rangle=0$ in which $|\Omega\rangle$ remains highest weight vector after duality transformation.

Sometimes fermionic duality transformation is sought as a particle-hole transformation for fermionic Bethe roots. This is not completely true as only part of dual fermionic roots are holes, see section~\ref{sec:stringglnm}.

\paragraph{Bosonic duality.} It is possible to perform duality transformation in the bosonic case also \cite{Pronko:1998xa,Gromov:2007ky,Bazhanov:2008yc}. It is based on the fact that Baxter equation allows for two solutions which are related by the Wronskian relation. Duality transformation exchanges these two solutions. In the compact case both solutions are polynomials and the pattern of the Bethe Ansatz equations after duality transformation does not change. In the noncompact case second solution is not a polynomial.

\subsection{\label{sec:rectangular}Rectangular representations}
For general choice of the highest weight representations the Bethe equations are not invariant under complex conjugation and therefore do not admit real solutions. The equations are however invariant if the weights satisfy one of the following two conditions\footnote{To be strict, invariance under complex conjugation may be achieved also if there are special relations between weights of different nodes of spin chain. There is an example in the literature \cite{Essler:2005ag} which fits into this class. Interestingly, solutions of Bethe Ansatz equations in \cite{Essler:2005ag} turned out to be not always real. We consider only spin chains with the same representation at each node.}:
\begin{subequations}
\label{possibleweights}
\be
      \label{pwa} {\rm either} && \hat m^{(\ell)}=-\hat m^{(\ell+1)}+c_\ell\\
      \label{pwb}  {\rm or}&&\hat m^{(\ell)}=\hat m^{(\ell+1)}.
\ee
\end{subequations}

Since we are aiming to formulate the string hypothesis, the reality of solutions is a necessary requirement, therefore we stick to the representations obeying (\ref{possibleweights}). Moreover, we will consider a simpler case of representations - those that  admit grading in which (\ref{pwa}) is true only for one instance of $\ell$. Such representations are characterized by the property that only one of the Dynkin labels is nonzero. On the language of  generalized Young tableaux inscribed into T-hook (see section~\ref{sec:unitary}) these are the representations given by a rectangle defined by position of its two corners: one at $\{0,0\}$,  another at $\{a,s\}\footnote{If $\{a,s\}$ corner of the rectangle belongs to right or left band of the T-hook, it defines the representation with more than one nonzero Dynkin labels. However, if we imbed the symmetry algebra in a larger one such that $\{a,s\}$ is in the vertical strip, it would be possible to choose grading where all except one Dynkin labels are zero. See appendix~\ref{sec:sourcewing} for more details.}$.  Due to this such representations are called rectangular.

Rectangular representations are all unitarizable, and as a consequence of unitarizability, their tensor power decomposes into irreps. For $\glnm$ case this is not a trivial property. In fact, appearance of indecomposables in tensor products of irreps in $\glnm$ is one of the reason that the results for $\gl(\groupn)$ case cannot be directly generalized. But rectangular representations allow for such generalization, and this  what is basically done in this paper.

Another appealing property of  rectangular representations is that they allow for a simple formulation of the unitary scattering matrix. Indeed, one can choose a spectral parameter in such a way that Lax operator satisfies $L(\theta)L(-\theta)\propto \MI$. The latter relation is a consequence\footnote{We thank to Carlo Meneghelli who pointed out this explanation to us.} of the property  $J_a^sJ_s^b\propto J_{a}^b$, where $J_a^b$ is the Lie algebra generator in a particular representation. The property $J_a^sJ_s^b\propto J_{a}^b$ holds for rectangular representations, but is not true in generic case.

\section{\label{sec:mainresult}String hypothesis and functional equations.}
\subsection{\label{sec:status}Status of string hypothesis.}
Let us explain a standard argument for appearance of string configuration on the example of the XXX magnet. The Bethe equations are written as
\be\label{twistedBetheXXX}
        \(\frac{u_i+\frac i2}{u_i-\frac i2}\)^L=-\prod_{j=1}^M\frac{u_i-u_j+i}{u_i-u_j-i}\,,
\ee
where $L$ is a length of the spin chain and $M$ is a number of Bethe roots.

The argument is the following. Suppose that $L$ is large and some Bethe root $u_n$ has a positive imaginary part: ${\rm Im}(u_n)>0$. Then the l.h.s of (\ref{twistedBetheXXX}) is exponentially large with $L$. To achieve this large value on the r.h.s. there should be another Bethe root $u_{n'}\sim u_n-i$, with the help of which the pole in the r.h.s. is created. Repeating the same arguments for $u_{n'}$ and using the reality of solution of the Bethe Ansatz \cite{VladimirovRealityBethe}, we conclude that the Bethe roots are organized in the complexes of the type:
\be\label{stringkkk}
       u_k= u_0+ik,\ \ \ k=-\frac{s-1}{2},-\frac{s-3}{2},\ldots,\frac{s-1}{2},
\ee
where $s$ is an integer. These complexes are called $s$-strings.

String hypothesis in its strong form states that all solutions of the Bethe Ansatz equations can be represented as a collection of strings, and that $u_k$ are approximated by $u_0+ik$ values with exponential in $L$ precision.

In its strong form the string hypothesis is wrong. However there is an evidence that its weaker  version is correct if the proper thermodynamic limit is taken. The weaker version states that most of the Bethe roots are organized into strings with exponential in $L$ precision, and that the fraction of solutions which significantly differ from (\ref{stringkkk}) decreases to $0$ when $L\to\infty$.    We discuss in more details applicability of the string hypothesis in appendix \ref{applicability}.

If the string hypothesis is valid at least in its weak form, and if energy of most of string-like excitations is low\footnote{this is so for the antiferromagnetic Hamiltonian.}, the approximation by strings is a good one. In the following we will work using this approximation. This allows us to rewrite the Bethe Ansatz equations into equations for centers of strings (called string Bethe Ansatz equations henceforth).

Further study, as it will be clear below, shows that approximation by string Bethe Ansatz equations is  good when Bethe Ansatz can be approximated by the linear integral equations for the densities of Bethe roots. We are going to derive now the functional equations equivalent to the linear integral equations, and the final result will be given by (\ref{mainequation}).

\subsection{Holomorphic projection.}
\subsubsection{$\su(2)$ case.}
In the large volume limit the Bethe Ansatz equations can be approximated by a functional equation. Let us show first how this is done for the solution of the $\su(2)$ spin chain (we are following \cite{Volin:2010cq} where more details can be found). The starting point is the Baxter equation
\be
        W^+Q^{--}+W^{-}Q^{++}=T\ Q,
\ee
where $W=u^L$, $Q=\prodl_{k=1}^M(u-u_k)$, and $T$ is the transfer matrix.

The ratio $r=\frac{W^+Q^{--}}{W^-Q^{++}}$  is assumed to have the property for $L\gg1$:
\be
|r|\gg1,&&{\rm Im}(u)>0\,,\no\\
|r|\ll1,&&{\rm Im}(u)<0\,.\label{rmagnitude}
\ee
This assumption is obviously not valid for large values of $u$. Note however that string hypothesis is also not valid at large values of $u$. For $u$ closer to origin ($u\lesssim\sqrt{L}$) assumption (\ref{rmagnitude}) is a good approximation. Moreover, below we will show how to derive exact string configurations from (\ref{rmagnitude}).

Based on (\ref{rmagnitude}), we can simplify Baxter equation to
\begin{subequations}
\be
  \label{holoup} W^+Q^{--}&=&T\ Q,\ \ \ {\rm Im}(u)>0\,,\\
  \label{holodown} W^-Q^{++}&=&T\ Q,\ \ \ {\rm Im}(u)<0\,.
\ee
\end{subequations}
We call this approximation the holomorphic projection of the Baxter equation.
We see that the l.h.s. of (\ref{holoup}) has zeroes at position $u_k+i$ where $u_k$ is a Bethe root.  R.h.s of (\ref{holoup}) should posses the same zeroes, therefore each such zero should be either zero of $Q$ or zero of $T$.

\paragraph{Accompanying roots.} Suppose first that all Bethe roots are real. Therefore $u_k+i$ cannot be zero of $Q$, thus it is a zero of $T$. Making the same consideration for equation (\ref{holodown}) we conclude that $T$ has also zeroes at $u_k-i$. We call $u_k\pm i$ accompanying zeroes. Therefore $T$ can be represented as
\be
        T=Q^{++}Q^{--}Q^*.
\ee
The zeroes of the polynomial $Q^*=\prod_{i=1}^{M^*}(u-u_{h,i})$ are called holes. All the holes should be real, otherwise $u_{h,i}\pm i$ will be the Bethe roots in virtue of (\ref{holoup}) and (\ref{holodown}).

By introducing the resolvents
\be\label{resolventdefinition}
        R=\pd_u\log Q,\ \  R^*=\pd_u\log Q^*
\ee
and taking the logarithmic derivative of (\ref{holoup}) and (\ref{holodown}) we get
\begin{subequations}
\be
        (1+D^{+2})R+R^*&=D^{+1}\frac{L}{u},&\ \ {\rm Im}(u)>0,\\
        (1+D^{-2})R+R^*&=D^{-1}\frac{L}{u},&\ \ {\rm Im}(u)<0.
\ee
\end{subequations}
These equations are equivalent to the linear integral equation which describe XXX spin chain in the thermodynamic limit (see introduction to part III of \cite{Volin:2010cq}).

\paragraph{Including string configurations.} If we allow the Bethe roots to be complex then the roots $u_k\pm i$ may be the Bethe roots as well. In theirs turn the Bethe roots $u_k\pm i$ may imply presence of $u_k\pm 2i$ roots e.t.c. In this way we
see that Bethe roots are organized in  string configurations once (\ref{rmagnitude}) is satisfied.

Let us rewrite the Baxter polynomial in the following form:
\be
        Q=\prod_{s=1}^\infty\CQ_{s}^{[s]_D},\   \CQ_s\equiv\prod_{k=1}^{\substack{M_s}}(u-u_{k,s}),
\ee
where $u_{k,s}$ is the center of $s$-string. Here we introduced the "$D$-number"
\be
[s]_D\equiv\frac{D^{s}-D^{-s}}{D-D^{-1}}\equiv D^{-s+1}+D^{-s+3}+\ldots+D^{s-1}
\ee
and used notation
\be
        Q^{\CF(D)}\equiv e^{\CF(D)\log Q}.
\ee

Repeating the logic of the accompanying roots, we conclude that the holomorphic projection of the Baxter equation leads to
\be\label{wfunctional}
        W^+=Q^*\prod_{s=1}^\infty \CQ_{s}^{D^{s+1}+D^{s-1}}.
\ee
The latter equation is written in the upper half plane. Equation in the lower half plane is just a complex conjugated one. We keep agreement to write only u.h.p. equations in the following.

The functional equation, the logarithmic derivative of (\ref{wfunctional}), is given by
\be\label{eqforstring1}
        (D+D^{-1})\sum_{s=1}^{\infty}D^s R_s+R_1^*=D\frac {L}u
\ee
with definition of resolvents analogous to (\ref{resolventdefinition}).

\paragraph{Functional equations for string configurations.} We obtained equation (\ref{eqforstring1}) which can be considered as equation for the density\footnote{Once the resolvent is known, the density is restored using (\ref{resoldef}).} of $1$-strings in the presence of higher-length strings. We will now derive equations for the densities of $s$-strings.

For this we should write Bethe Ansatz equations for the center of strings. They are obtained by multiplication of the Bethe Ansatz equations for the Bethe roots - centers of strings. The result is the following

\be\label{Bethestrings}
        \CF_s(u_{k,s})=1,
\ee
where we defined
\be
\ \ \ (-1)^s\CF_s(u)\equiv W^{D^s-D^{-s}}\prod_{s'=1}^\infty \CQ_{s'}^{(D+D^{-1})\CL_{ss'}},\ \ee
\be
 \CL_{ss'}\equiv\frac{(D^s-D^{-s})(D^{s'}-D^{-s'})}{D-D^{-1}}\ .
\ee
Then we introduce a Baxter-like equation, perform its holomorphic projection and use the argument for accompanying roots to get functional equations. The net result for the rules to obtain functional equations is the following:
\bn
        \item Write Bethe Ansatz equation in the form
        \be
                \CF(u_i)=1, \ \ \CF(u)=\prod_k\frac{{f_k}^{\CP(D^{+1})}}{f_k^{\CP(D^{-1})}}\,,
        \ee
        where $f_k$ is some function of $u$ (in the case of (\ref{Bethestrings}) this is $W$ or $\CQ_s$), and $\CP(t)$ is a polynomial (or more generally, series) in $t$ such that $\CP(0)=0$.
\item The functional equations for the resolvents $R_k=\pd_u\log f_k$ and $R=\sum\frac 1{u-u_i}$ are then written in the form:
\be
        R+R^*=\pm \sum_{k}\CP(D^{+1})R_k,\ \ {\rm Im}(u)>0,
\ee
and the conjugated equation in the l.h.p.
\item The sign is chosen to be positive if $|\CF(u)|\gg1$ for ${\rm Im}(u)>0$ and to be negative if $|\CF(u)|\ll 1$ for ${\rm Im}(u)>0$.
\en

Application of these rules to (\ref{Bethestrings}) gives:
\be
        R_s^*+(D+D^{-1})\sum_{s'=1}^\infty \CL_{ss'}^+R_{s'}=D^s\frac{L}{u}\,.
\ee
Now note that $(\CL^+)^{-1}_{ss'}=C_{ss'}$, where
\be
        C_{ss'}=(D+D^{-1})\delta_{ss'}-\delta_{s,s'+1}-\delta_{s,s'-1},\ \ \ s,s'\geq 1.
\ee
Therefore we get
\be{\label{vcool}}
    (D+D^{-1})R_s+\sum_{s'=1}^\infty C_{ss'}R_s^*=\delta_{s,1}\frac Lu.
\ee

\subsubsection{$\su(\groupn)$ case.}
Exactly the same procedure can be applied for the Bethe equations in the $\su(\groupn)$ case, and in general any compact rectangular representation $\{A,S\}$ and the case when inhomogeneities $\inh_k$ are present can be considered. The result is
\be\label{vvcool}
        \sum_{a'=1}^{\groupn-1}C_{aa'}R_{a',s}+\sum_{s'=1}^\infty C_{ss'}R_{a,s'}^*=\delta_{a,A}\delta_{s,S}J\,,
\ee
where $R_{a,s}$ is the resolvent for the density of $s$-strings composed from the Bethe roots that corresponds to the simple root $\a_a$ of the Dynkin diagram. $J$ stands for the resolvent of a source term and is given by  $J=\prodl_{k=1}^{L}\frac 1{u-\inh_k}$.

If we take the difference of (\ref{vvcool}), defined for $\rm{Im}(u)>1/2$, and the complex conjugation of (\ref{vvcool}), defined for ${\rm Im}(u)<1/2$, we get exactly the equation (\ref{mainequation}). To define this difference,  a prescription for analytical continuation is needed, but this is also the case for (\ref{mainequation}).

\subsection{\label{sec:stringglnm}String hypothesis for $\glnm$ case.}

In the supersymmetric case a new type of the Bethe equations is added - equation for the fermionic node. Assume the following labeling of the Bethe roots:

\be\label{uthvlabeling}
    \raisebox{0.1cm}{\parbox[c]{5cm}{\includegraphics[width=5cm]{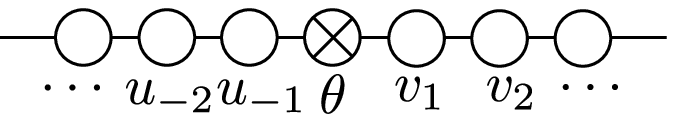}}}\ \ ,
\ee
and for simplicity $u_{-1}=u$, $v_1=v$.

The number of Bethe roots decreases from left to right: $\ldots>M_{u}>M_{\theta}>M_{v}>\ldots$; this ordering defines what term in the Baxter equations should be suppressed for the holomorphic projection.

The Bethe equations for the $\theta$-roots are written as
\be
        1=\prod_{k=1}^{M_{u}}\frac{\theta_i-u_k+\frac i2}{\theta_i-u_k-\frac i2}\prod_{k=1}
^{M_{v}}\frac{\theta_i-v_k-\frac i2}{\theta_i-v_k+\frac i2}.
\ee
The QQ-relation is the the following:\be
Q_u^+Q_v^--Q_u^-Q_v^+=Q_\theta\overline Q_\theta.
\ee
Here $Q_u,Q_v,Q_\theta$ stand for the Baxter polynomials of the corresponding Bethe roots and $\overline Q_\theta$ is for dual fermionic roots, see section \ref{sec:duality}.

The holomorphic projection gives
\be\label{holofermion}
\begin{array}{ccc}
      Q_u^+Q_v^- &=& + Q_\theta\overline Q_\theta,\ \  {\rm Im}(u)>0\,,\\
      Q_u^-Q_v^+ &=& - Q_\theta\overline Q_\theta,\ \  {\rm Im}(u)<0\,.
\end{array}
\ee

We see that since there is no $Q_\theta$ term on the l.h.s. of the QQ-relation, fermionic Bethe roots cannot form string configurations by themselves. Suppose however that among the $u$-roots there is a string of length $s$ with center at $u_0$. Holomorphic projection (\ref{holofermion}) implies then that there is a string of the length $s-1$ (with the same $u_0$ center) which should be the string\footnote{The reader may ask why the string cannot be distributed among both $Q_\theta$ and $\overline Q_\theta$. The reason is in a back-reaction of $\theta$ Bethe roots in the Bethe equations for $u$. This is most clearly seen through the duality transformations.} of $Q_\theta$ or $\overline Q_{\theta}$. Equivalently, presence of $s$-string among $v$-roots implies presence of $(s+1)$-string either in $Q_\theta$ or $\overline Q_\theta$.

\subsubsection{Triangles (distinguished diagram).}Let us focus on the case of $s$-string of $v$-roots. Suppose that induced $(s+1)$-string is that of $\overline Q_\theta$. Then, since $\overline\theta$ roots do not interact with $v$ roots, there is no back reaction of $(s+1)$-string, and $s$-string of $v$-roots is just a string configuration discussed in the bosonic case. Situation is however different if $(s+1)$-string is the one of $Q_\theta$. Let us perform the duality transformation on the $\theta$-node. Then $v$-node becomes fermionic, and it still contains the $s$-string. Fermionic node cannot contain $s$-string by itself, and there is also no $(s+1)$-string of $\bar\theta$ that would induce  the $s$-string of $v$-s. Therefore there should exist an $(s-1)$-string of $v_2$. The procedure of dualisation is applied again, now to the node $v_1$, and so forth until we reach the $1$-string at the node $v_{s-1}$. Coming back to the variables before dualisations, we conclude that an $(s+1)$-string of fermionic Bethe roots may exist if it is a part of the following "triangle" configuration:
\be\label{dynkintriangle}
        \parbox{3.5cm}{\includegraphics[width=3.5cm]{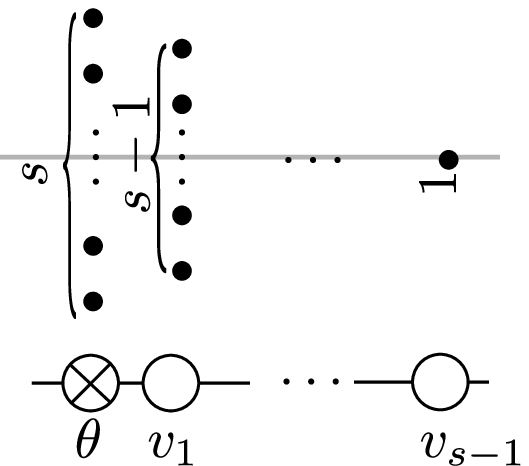}}.
\ee
We will call this configuration $(s+1)$-triangle.
\subsubsection{Trapezia (non-distinguished diagram).}
The trick with fermionic duality transformation can be applied to reveal another possible configuration. Let us come back to the case when there is an $s$-string of $v$-s and an induced $(s+1)$-string is that of $\overline\theta$. By performing the duality transformation on the $\theta$ node we get the so called trapezium configuration which consists of $s$-string of $v$-s and $(s+1)$ string of $\bar\theta$. Note that $v$ becomes a fermionic node after the duality transformation.

The general configuration called $(s,s')$-trapezium is described as follows:
\be
        \parbox{3.5cm}{\includegraphics[width=3.5cm]{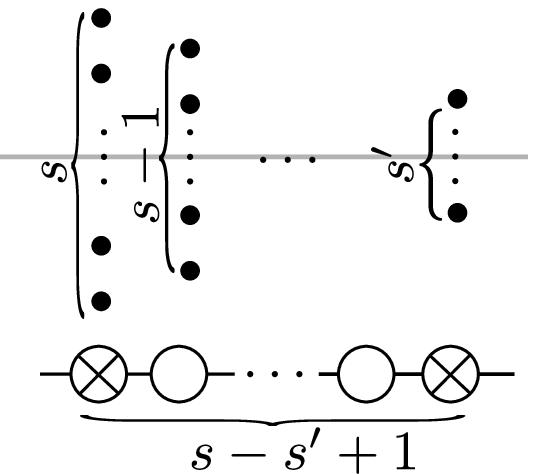}}.
\ee
This configuration can be obtained after for example by the set of duality transformations $d1$ or $d2$:
\be\label{dualitytricks}
        \parbox{10cm}{\includegraphics[width=10cm]{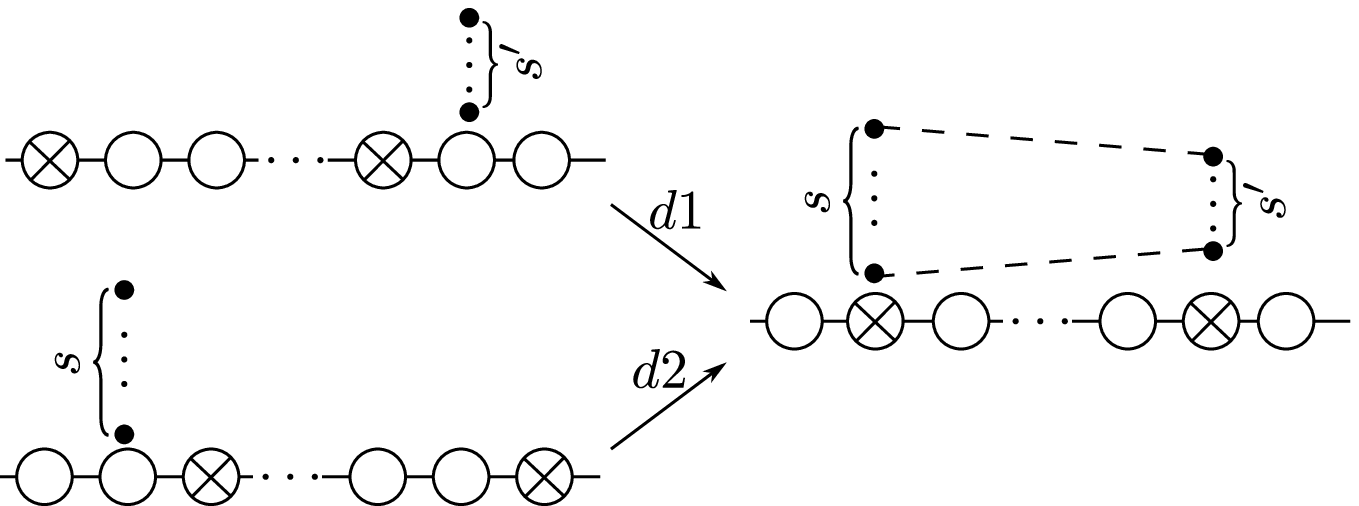}}.
\ee

The remarks follow:
\bi
        \item[-] There are some complications  if the duality transformation is applied to the node with the source. In general situation, fermionic nodes - boundaries of trapezium cannot be separated by the node with the source.
        \item[-] Any trapezium becomes a bosonic string for a certain choice of Kac-Dynkin diagram. To see this, we just make duality transformations $d1$ or $d2$  back.
We have a choice to obtain either $s$ or $s'$ string.
        \item[-] Triangle can be always seen as unfinished trapezia. For this we  need to enlarge the Dynkin diagram (\ref{dynkintriangle}) with one auxiliary (virtual) fermionic node. The enlargement can be always done, with the additional requirement that there is no Bethe roots at the virtual node. After this trick, we can make duality transformations similar to the inverse of $d2$ to bring the $s$-triangle to the $s$-string. However, virtual fermionic node will be dualized, and we should require maximal excitation at the dualized virtual node.

\item[-] Duality tricks\footnote{Somehow similar tricks were used in \cite{Freyhult:2009fc} to find a new class of explicit solutions of 1-loop Bethe Ansatz for AdS/CFT spectral problem.} allow us to make even a more general transformation. An $s$-string at node $k$ for one choice of the Kac-Dynkin diagram can be represented as an $s'$-string at node $k'$ for some other choice of the diagram if $s-k=s'-k'$. To perform this transform we may probably introduce a virtual fermionic node.
\ei

\subsubsection{Diamonds (noncompact, both $\gl(\groupn)$ and $\glnm$).}
Consider now noncompact case and the diagram of the form:
\be
    \raisebox{0.1cm}{\parbox[c]{4cm}{\includegraphics[width=4cm]{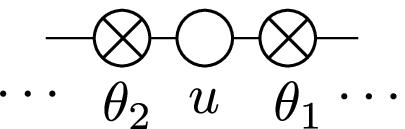}}}\ ,
\ee
where $u$ is a momentum-carrying node.
If the nonzero Dynkin label $m$ is positive, the Bethe roots of type $u$ may form $s$-strings. After dualisation of $\theta_1$ and $\theta_2$ nodes the $u$ node transforms to noncompact one and the $s$-strings become a set of two trapezia. This shape will be called diamond. The general $(s,s',s'')$-diamond configuration looks as follows:
\be
    \raisebox{0.0cm}{\parbox[c]{4cm}{\includegraphics[width=4cm]{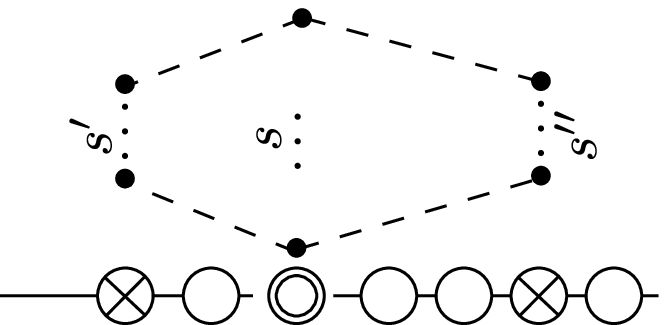}}}\ ,
\ee

$(s,s',1)$-diamond can terminate also on the bosonic nodes, as triangles do. In particular, $(s,1,1)$ configurations can exist in a pure bosonic noncompact case.

Diamonds can be brought to the bosonic string form with the help of fermionic duality transformations. In the $\su(\groupn,\groupm)$ case this can be done only after introducing of a virtual fermionic node.

\subsection{\label{sec:functionalglnm}Functional equations for $\glnm$ spin chain.}
After we gave a description of string configurations that appear in $\glnm$ spin chains, it's a time to write down functional equations that involve resolvents for densities of these configurations. We will consider first compact $\su(\groupn|\groupm)$ case and distinguished Kac-Dynkin diagram. The generalizations then can be easily done.

 The distinguished Kac-Dynkin diagram allows, apart usual bosonic strings, only for triangle configurations. Let us  understand how triangles enter to the string Bethe equations for the other nodes. The node $u_{-1}$ (in the notation of (\ref{uthvlabeling})) adjacent to the fermionic base of $s$-triangle interacts with triangle as with $s$-string. The node $v_s$ interacts with triangle as with real Bethe root. Bosonic nodes $v_k$, $1\leq k<s$, do not interact with triangle. Indeed, Bethe equation for these bosonic nodes is written as
\be\label{Bethevk}
        -1=\frac{Q_{v_{k-1}}^+}{Q_{v_{k-1}}^-}\frac{Q_{v_{k+1}}^+}{Q_{v_{k+1}}^-}\frac{Q_{v_{k}}^{--}}{Q_{v_{k}}^{++}}.
\ee
let us represent $Q_{v_k}=Q'_{v_k}\prodl_{s>k}\CQ_{\Delta,s}^{[s-k]_D}$, where $\CQ_{\Delta,s}$ - the Baxter polynomial for the centers of $s$-triangles and $Q'$ - the Baxter polynomial for all other Bethe roots. It is easy to see that all $\CQ_{\Delta,s}$ cancels out from (\ref{Bethevk}).

Procedure of holomorphic projection and introduction of accompanying roots for bosonic Bethe roots remains the same as previously. It leads to the following functional equations for all purely bosonic string configurations:
\be\label{vcoolll}
 C_{a,a'}R_{a',s}+C_{s,s'}R_{s,s'}^*={\rm source\ term}.
\ee
This equation has the same form as (\ref{vvcool}), but identification of the resolvents with string configurations is slightly more involved. It should be done in the following way. First, assign string configurations to the nodes of the two-dimensional lattice as shown in the figure:\be
       \raisebox{1.2cm}{\parbox{5cm}{\includegraphics[width=5cm]{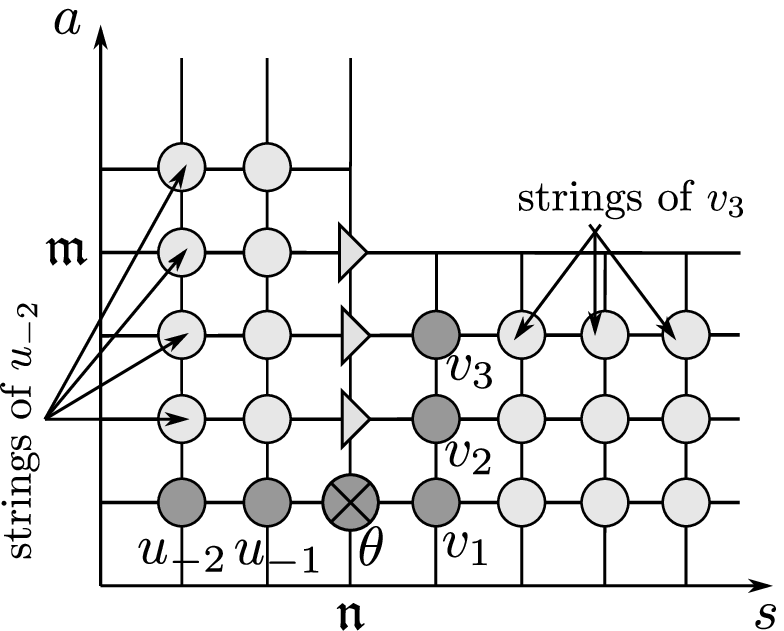}}}\ .
\ee
String configurations which include fermionic Bethe roots are situated on the line $s=\groupn$. All other nodes correspond to bosonic strings as shown on the figure. If $s>\groupn$, $R_{a,s}$ is identified with the resolvent for string configurations at $\{a,s\}$ node and $R_{a,s}^*$ -- with the resolvent for correspondent holes. For the case $s<\groupn$ holes and particles are interchanged: $R_{a,s}$ is defined to be the resolvent of holes, while $R_{a,s}^*$ is defined to be the resolvent for particles.

With the assignments for the resolvents as above, (\ref{vcoolll}) is certainly valid for arbitrary $a$ and $s\neq\groupn$.
It is left to show that the same equation holds if $s=\groupn$, i.e. when it comes from the  functional equations for triangles. The simplest way to do this is to add the virtual fermionic node and to dualize the Kac-Dynkin diagram using inverse of $d2$ in (\ref{dualitytricks}). The $s$-triangles become bosonic $s$-strings, and therefore the resolvents for them should obey the same equation (\ref{vcoolll}).

\begin{figure}[t]
\centering
\includegraphics[height=4cm]{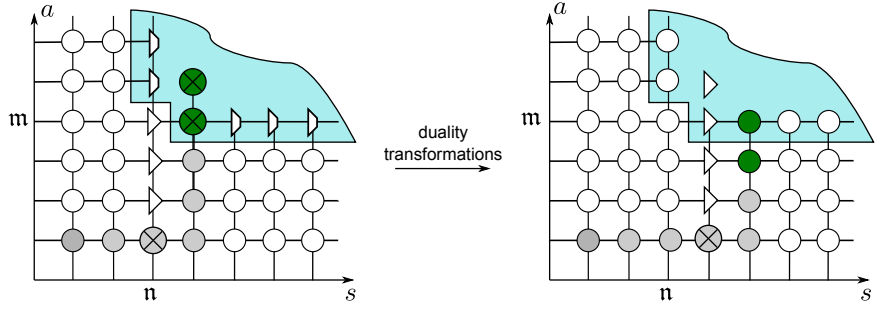}
\caption{\label{fig:cornerequation}\footnotesize Figure showing how to derive functional equation for the triangle configurations and for the corner. Gray circles show position of real Bethe roots for original Kac-Dynkin diagram. Green circles show two auxiliary virtual Bethe roots. Blue region is the place where excitations are derived with the help of virtual Bethe roots and therefore not physical. Thus we impose in the blue region $R_{a,\groupn}^*=0$, $R_{\groupm,s}=0$ and $R_{\groupm+1,\groupn+1}=R_{\groupm+1,\groupn+1}^*=0$.
We see that after duality transformation triangles on the line $s=\groupn$ transform into strings and therefore should obey standard functional equations of the form (\ref{vcoolll}). Equation (\ref{vcoolll}) in the corner $\{\groupm,\groupn\}$ includes $R_{\groupm+1,\groupn}$ and $R_{\groupm,\groupn+1}^*$ and therefore not explicitly expressed through the physical objects. }
\end{figure}

The only situation when introduction of virtual fermionic node does not directly work, is the longest possible triangle. Indeed, this triangle is situated at position $\{\groupm,\groupn\}$. It interacts with dualized virtual fermionic node and this interaction is not expressed through the physical objects. To express equation in the corner only through the physical resolvents, let us consider the following sets of equations (for $a,s\geq 1$):
\be
        C_{aa'}R_{\groupm-1+a',\groupn-1+s}+C_{ss'}R_{\groupm-1+a,\groupn-1+s'}=\d_{a,1}R_{\groupm-1,\groupn-1+s}+\d_{s,1}R^*_{\groupm-1+a,\groupn-1}.
\ee
Now, act on these equations with $\suml_{a,s=1}^\infty (\d_{s,1}D^{a}+\d_{a,1}D^{s})$, the result is
\be\label{cornerequation}
R_{\groupm,\groupn}+R_{\groupm,\groupn}^*=\sum_{s=1}^{\infty}D^{s}\left(R_{\groupm-1,\groupn-1+s}+R_{\groupm-1+s,\groupn-1}^{*}\right)\,.
\ee
This nonlocal equation is expressed only through the physical resolvents and this is the equation that should be considered in the corner.

Any dualisations of Dynkin diagram will result in reinterpretation of strings as triangles and trapezia and vice versa. However, the main equation (\ref{vcoolll}) remains unchanged, once this reinterpretation is taken into account.

We are ready to formulate the main statement of this section.

\paragraph{Proposition.}\hspace{-1.3EM}\footnote{Except for the equation (\ref{cornerequation}) for the corner this proposition was initially formulated in \cite{Volin:2010cq}. Here we add statement about the corner equation and significantly simplify derivation of (\ref{onesimpleformofmaineequation}) based on duality transformations.} \textit{Consider some Kac-Dynkin diagram for a given} $\glnm$ \textit{algebra and rational integrable spin chain of length} $L$ \textit{and inhomogeneities} $\inh_1,\ldots \inh_L$. \textit{Suppose that all the nodes of spin chain are in the same highest weight irrep with respect to chosen Kac-Dynkin diagram. If this irrep is such that all Dynkin labels are} $0$ \textit{except one which equals to} $m$\textit{, then it is possible to formulate string hypothesis. In the thermodynamic limit Bethe Ansatz equations are approximated by the  functional equations of the following form:}
\be\label{onesimpleformofmaineequation}
        C_{aa'}R_{a',s}+C_{ss'}R_{a,s'}^*=\delta_{s,1}\delta_{a,|m|}J,\ \ J=\sum_{i=1}^L\frac 1{u-\inh_k}.
\ee
\textit{This equation is most generally valid in the domain called $T$-hook which is explicitly constructed below. In the corners of the $T$-hook $R_{a,s}$ and $R_{a,s}^*$ satisfy equation (\ref{cornerequation})}.

Construction of $T$-hook and identification of $R_{a,s}$ and $R_{a,s}^*$ with resolvents of string configurations is done by the following procedure:
\bn
        \item First draw Kac-Dynkin diagram on the two-dimensional square lattice by the following rule. Put the node with nonzero Dynkin label at position $a=1,\ s=0$. Then for the nodes to the right start moving right and change direction to up or right at each node next to the fermionic\footnote{Let us stress that the rule we apply is different from the one in \cite{Zabrodin:2007rq}, where the path of Backlund transformations was encoded by Dynkin diagram which turns at each fermionic node. We  also apply the rule to turn at the fermionic node, for instance in Fig.~\ref{fig:duality}, but when we consider the questions of grading, not organization of string configurations.}. For the nodes to the left start moving left and change direction to up or left at each node next to the fermionic.
 Note that the source node can be bosonic, right or left fermionic or double fermionic (noncompact).
\item Configurations of real Bethe roots are at the position occupied by the nodes of the Dynkin diagram.

\item String configurations with the base node $\Upsilon$\footnote{For simple strings $\Upsilon$ is just the type of Bethe roots from which strings are made. For trapezia, triangles, and diamonds $\Upsilon$ is the node containing the longest string.} are on the ray that originates from the corresponding Dynkin node $\Upsilon$ and goes perpendicularly  to the link between $\Upsilon$ and adjacent Dynkin node with smaller number of the Bethe roots.

\item If the lines containing string configurations go horizontally, then $R_{a,s}$ is the resolvent of density of the strings at node $\{a,s\}$ and $R_{a,s}^*$ is the resolvent for hole density. If the lines go vertically, then $R_{a,s}^*$ is the resolvent for string density and $R_{a,s}$ is the resolvent for hole density.
\en

For main examples of this construction we refer to Fig.~\ref{fig:patterns}. Below we add one more sophisticated example, where with blue line we marked imbedding of the Kac-Dynkin diagram and with red lines we marked collection of string configurations with the same base:

\be\label{onemoreexample}
        \parbox{5.1cm}{\includegraphics[width=5.1cm]{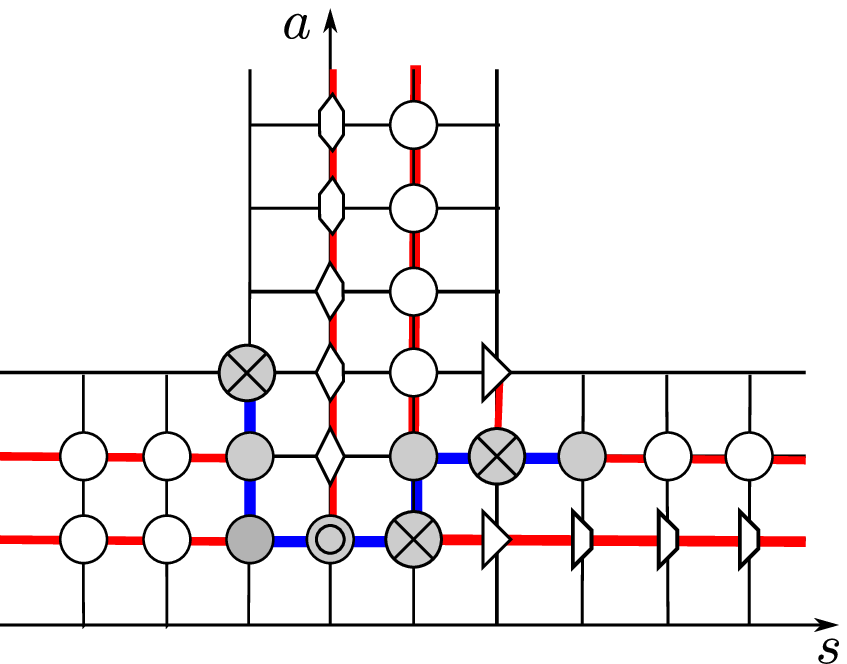}}
\ee

\section{\label{sec:unitary}T-hook and {\it all} unitary highest weight representations}

In \cite{Gromov:2010vb} Gromov, Kazakov, and Tsuboi proposed to inscribe a generalization of  Young tableaux into arbitrary T-hook and gave a rule how to define the unitary highest weight representation (UHW) from a given inscription. This rule is distinguished in a sense that characters of representations
assigned to rectangular Young tableaux
satisfy bilinear identity
\be\label{charhirota}
\chi_{a,s}^2=\chi_{a,s+1}\chi_{a,s-1}+\chi_{a+1,s}\chi_{a-1,s},
\ee
where characters are nontrivial for $\{a,s\}$ inside the T-hook domain, equal to $1$ on the boundaries, and zero outside.

Quantum generalization of (\ref{charhirota}) is the Hirota equation on the transfer matrices.

\ \\
Here we would like to make a conjecture that the map of \cite{Gromov:2010vb} from generalized Young tableaux to UHW is in fact a bijection.

The proof of the conjecture is just to take the mapping rule in \cite{Gromov:2010vb} and to compare it       with existing classification theorems of UHW in the literature. For $\su(\groupn)$, $\su(\groupn,\groupm)$, $\su(\groupn|\groupm)$, and $\su(\groupn,|\groupm)$ real forms such proof goes smoothly. Unfortunately, for the most interesting $\su(\groupn,|\groupm\,|\,\groupk)$ case we found an apparent contradiction in the literature (see discussion below) which prevents for the moment to put the conjecture in a form of the theorem.\com{ From \cite{JakobUHW} it follows that all UHW can be represented in the way done in \cite{Gromov:2010vb}, however there are some Young tableaux which give nonunitary representations. From classification by Dobrev and Petkova \cite{Dobrev:1985qv} of UHW for $SU(2,2|\groupn)$ it follows that our statement about bijection is correct, and one can check that classification of \cite{JakobUHW} applied to $SU(2,2|\groupn)$ case is in contradiction with the results of  \cite{Dobrev:1985qv}. Therefore
our statement about bijection remains a plausible
conjecture before this contradiction is resolved.} Below we will explain what the bijection conjecture should imply and for what class of representations the bijection conjecture still has to be clarified.

Before giving the list of UHW according to the bijection conjecture, we first recall the definition of real forms and unitary representations.

\paragraph{Real form.} Consider the Lie algebra $\cg$ over $\MC$. A real form of $\cg$ is by definition a such algebra $\cg_0$ over $\MR$ that
\be
    \cg=\cg_0\otimes_{\MR}\MC.
\ee
A possible way to define the real form is to introduce
an involutive antilinear automorphism $^*$. It is straightforward to check that generators $J_0$ of the algebra that satisfy $J_0^*=-J_0$ can be used to span $\cg_0$.

If instead $\cg_0$ is given, one can define $^*$ by the property $J_0^*=-J_0$ and $(a\, J)^*=\overline a J^*$, where $a$ is a complex number, $J_0\in \cg_0$, $J\in \cg$, and $\overline a$ means complex conjugation.

\paragraph{Unitary representation.}
The representation $T$ is called unitary for a given real form if we can define a hermitian form (positive definite scalar product) on the representation module such that for all $J\in \cg$
\be
    T(J^*)=T(J)^+,
\ee
where $^+$ means hermitian conjugation with respect to the hermitian form.

\paragraph{Possible real forms.}We will focus on the real forms that allow to choose a Borel decomposition such that
\be
        E_{ii}^*=E_{ii},\ \  E_{ij}^*=(-1)^{c(i)+c(j)}E_{ji},
\ee
where $c(i)=0\ {\rm or}\ 1$; this is a $\MZ_2$ grading different in general from the grading $|i|$. The root corresponding to $E_{ij}$ is called compact if $c(i)+c(j)$  is even and noncompact if $c(i)+c(j)$ is odd.

\begin{figure}[t]
\begin{centering}
\begin{tabular}{|c|c|c|c|c|c|}
\hline
\parbox{4em}{\ \ \includegraphics[width=2.7em]{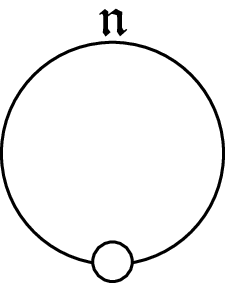}} &
\parbox{4em}{\includegraphics[width=3.9em]{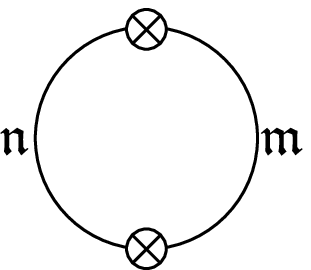}} &
\parbox{4em}{\includegraphics[width=3.9em]{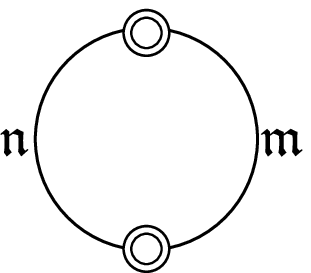}} &
\parbox{3.9em}{\includegraphics[width=3.9em]{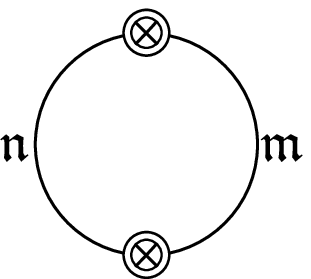}} &
\parbox{3.9em}{\includegraphics[width=3.9em]{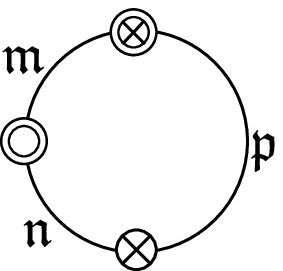}} &
\parbox{3.9em}{\includegraphics[width=3.9em]{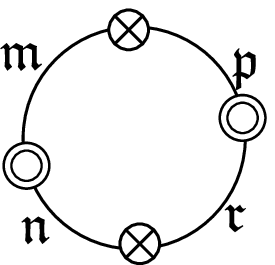}} \\\hline
\!$\su(\groupn)$\! & \!$\su(\groupn|\groupm)$\! & \!$\su(\groupn,\groupm)$\! &  \!$\su(\groupn,\!|\groupm)$\! & \!$\su(\groupn,\groupm|\groupk)$\! & \!$\su(\groupn,\groupm|\groupk,\groupr)$\! \\ \hline
\end{tabular}
\caption{\label{fig:possiblerealform}List of possible real forms. Cross in a Dynkin node means fermionic root, as before. Additional circle around a Dynkin node means noncompact root. Number $\groupn$  over the line means $\groupn-1$ compact bosonic roots, the same for $\groupm,\groupk,\groupr$.}
\end{centering}
\end{figure}

Let us consider set of positive simple roots (that correspond to $E_{i-1,i}$) together with the negative one corresponding to $E_{\groupn+\groupm,1}$. In this set the number of noncompact roots is always even, the same is true for the number of fermionic roots. By performing fermionic duality transformations we can always reduce number of compact and fermionic simple roots to at most $2$. As a result the full set of real forms and corresponding distinguished gradings is summarized in Fig.~\ref{fig:possiblerealform}.

In the following we will consider UHW representations with respect to the standard Borel subalgebra defined by one of the cases in Fig.~\ref{fig:possiblerealform}\footnote{There are other choices of Borel subalgebra which lead to a different class of highest weight representations. For example, Dynkin diagram of three noncompact bosonic nodes still corresponds to the case of $\su(2,2)$ real form. Highest weight irreps in this case are not always equivalent to highest or lowest weight irreps defined by Dynkin diagram with standard compact-noncompact-compact signature. For some reason such alternative choices of Borel subalgebra were not systematically studied in the known to the author literature, though they give new instances of UHW \cite{characterformula}. A class of Bethe Ansatz equations we consider corresponds to a standard choice of Borel subalgebra, thus we focus on UHW with respect to this standard choice.}.

\paragraph{UHW for $\su(\groupn)$.} In this case string hypothesis lead to a strip of width $\groupn$ (case (a) in Fig.~\ref{fig:patterns}). We inscribe a standard Young tableaux in this strip which defines the UHW with weight $[m_1,\ldots,m_\groupn]$:
\be
    \raisebox{0.4cm}{\parbox[c]{5.5cm}{\includegraphics[width=5.5cm]{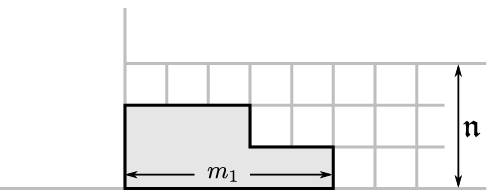}}}\  \ .
\ee

\paragraph{UHW for $\su(\groupn,\groupm)$.}  The maximal compact subalgebra is $\su(\groupn)\oplus \su(\groupm)\oplus \ualgebra(1)$, with the $\ualgebra(1)$ generator defined by:
\be
        E_{\ualgebra(1)}=\frac 1{\groupn}\sum_{k=1}^{\groupn} E_{kk}-\frac 1{\groupm}\sum_{k=1}^{\groupm} E_{kk}.
\ee
We denote the weight by
\be
        [-a_1+\l,-a_2+\l,\dots,-a_\groupn+\lambda,c_1,c_2,\ldots,c_\groupm]
\ee
with condition $a_\groupn\geq a_{\groupn-1}\geq \ldots \geq a_1=0$ and $c_1\geq c_2\ldots\geq c_\groupm=0$. The necessary condition to get unitarity is that $[a_\groupn,\ldots,a_1]$ and $[c_1,\ldots,c_\groupm]$ define unitary representations for $\su(\groupn)$ and $\su(\groupm)$ subalgebras respectively. The $a-$ and $c-$ Young tableaux are inscribed into a slim hook as follows\footnote{The turning points of the partitions are at the nodes of the lattice, they are shifted from the nodes only to improve readability of the picture.}

\be\label{suknkhok}
    \raisebox{0.7cm}{\parbox[c]{7.5cm}{\includegraphics[width=7cm]{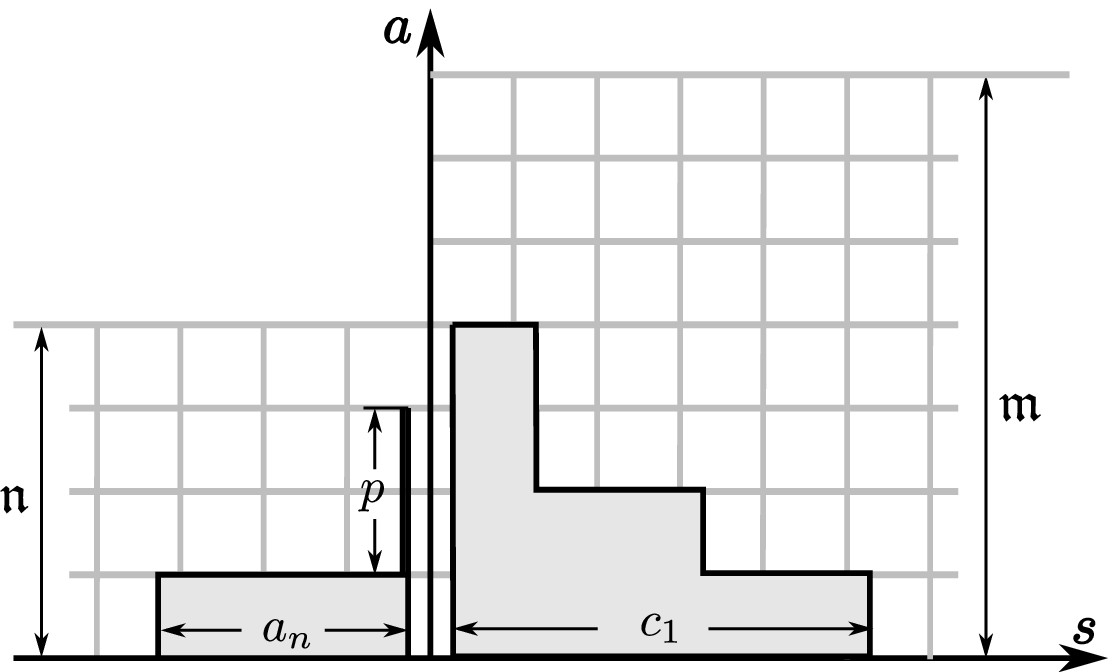}}}.
\ee
To the $a$-Young tableaux we added a vertical line of length $p$ which allows map value of $\l$ as
\be
        \l=-p-p_a-p_c,
\ee
where $p_a$($p_c$) is the number of nonzero $a_k$($c_k$). So, $\l$ is minus the total height of the figure inscribed into the slim hook. Value of $p$ should be a nonnegative integer if $p\leq p_{0}={\rm Max}(\groupn-p_a-1,\groupm-p_c-1)$ and can be arbitrary real if  $p\geq p_0$.

Comparison with the classification theorem in \cite{MR733809} shows that in a described way we parameterize exactly all UHW for the $\su(\groupn,\groupm)$ real form and standard choice of Borel decomposition.

If we also allow add vertical line to the $c$-tableaux and define $p$ as a total length of $a$ and $c$ vertical lines then restrictions on $p$ are quite natural from the graphical point of view. Indeed, if the figure is completely inside the slim hook, it should belong to the integer lattice and the corresponding representation is a short one, analog of the BPS state. If the figure can be made, by assigning $p$ only to $a$ or $c$, such that the distance to the boundary is less then one, representation becomes long and $p$ can be deformed continuously.

There is also a nice interpretation of $p$ from the point of view of oscillatory description. To construct $\su(\groupn)$ representation with Young tableaux of height $p_a$, one need at least $p_a$ copies of $n$ oscillators to be able performing antisymmetrization. Nonzero integer $p$ means that we take more copies of oscillators than needed, namely $p_a+p$.
\ \\

\quad
\paragraph{UHW for $\su(\groupn|\groupm)$.} This case corresponds to covariant $\glnm$ representations -- the ones that can be obtained from tensor power of the fundamental representation. All UHW  are bijected to the Young tableaux that can be inscribed into a fat hook:
\be\label{suknmhok}
    \raisebox{0.7cm}{\parbox[c]{6cm}{\includegraphics[width=5.5cm]{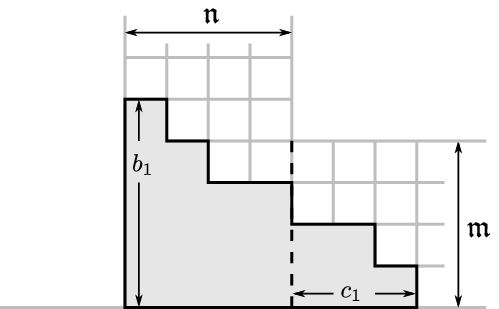}}}\ \ \ .
\ee
The weight of the representation is given by
\be
        [b_1,\ldots,b_{\groupn},c_1,\ldots,c_{\groupm}],
\ee
Note that $b_\groupn$ plays the same role as $p+p_c$ in the $\su(\groupn,\groupm)$ case. Restrictions on $b_\groupn$ to get UHW are: first, $b_\groupn\geq p_c$; second, $b_\groupn$ should be integer if $b_\groupn<\groupm-1$ or can be arbitrary real number otherwise. If $b_\groupn\geq \groupm$, one can put $c_\groupm$ to arbitrary nonnegative real number.

\paragraph{UHW for $\su(\groupn,|\groupm)$.}  This case corresponds to contravariant representations of $\glnm$ - the ones that can be obtained from tensor power of the antifundamental representation. Though we formally have noncompact generators $E_{ij}$ in a sense that $c(i)+c(j)=1$, all such generators are fermionic and therefore the UHW is finite dimensional. The classification rules are the same as in the previous case, just the role of $n$ and $m$ is exchanged and assignment of the weights goes as follows:
\be\label{suknmhok}
    \raisebox{0.7cm}{\parbox[c]{5.5cm}{\includegraphics[width=5.5cm]{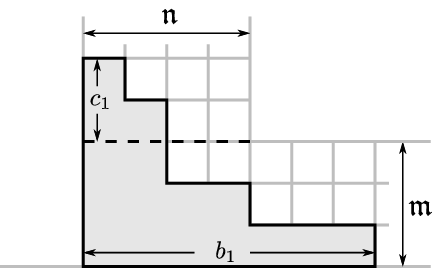}}}\ \hspace{1em},\hspace{2em}\\
\no \\
       {\rm highest\ weight\ is\ \ \ } [-c_\groupn,\ldots,-c_1,-b_\groupm,\ldots,-b_1].
\ee

\paragraph{UHW for $\su(\groupn,\!|\groupm|\groupk)$}  By fermionic duality transformations the $\su(\groupn,\!|\groupm|\groupk)$ case can be mapped to the case $\su(\groupn,\groupk|\groupm)$ - the one for the distinguished Dynkin diagram, so we won't consider the $\su(\groupn,\groupk|\groupm)$ case  separately. We choose a grading in which the first basis vector $v_i$ is odd: $|1|=1$. The parameterization of the highest weight representations is the same (up to some relabeling) as in \cite{Gromov:2010vb}:

\be
\raisebox{2.7em}{\parbox{0.55\textwidth}{\includegraphics[width=0.5\textwidth]{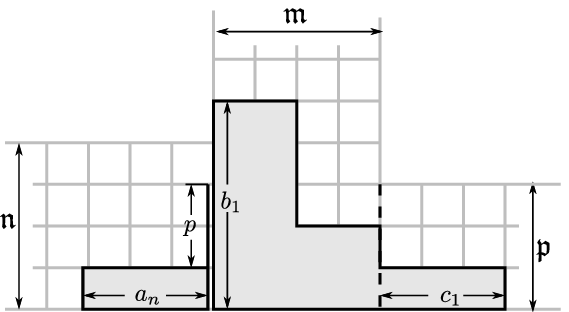}}}\ .
\ee
The highest weight is given by
\be
        [\l,-a_2+\l,\ldots,-a_\groupn+\l,b_1,\ldots b_\groupm,c_1,\ldots c_\groupk]
\ee
with $\l=-p-p_a-b_1$ (again minus the total height of the inscription). Furthermore, one requirers that $p\geq 0$,  $b_\groupm\geq p_c$ and that $p$ and $b_\groupm$ are integers if respectively $p+p_a\leq \groupn-1$ or $b_\groupm\leq k-1$ (and this corresponds to BPS states).  $p$ ($b_\groupm$) can be continuous if $p+p_a> \groupn-1$ ($b_\groupm> k-1$).
Let us translate these requirements to the possible values of the fermionic Dynkin labels\footnote{here $n,m,k$ mean number plus one of the compact bosonic nodes on the corresponding node.} $\hat r_1$ and $\hat r_2$,
\be
\raisebox{0.3cm}{\parbox{0.35\textwidth}{\includegraphics[width=0.35\textwidth]{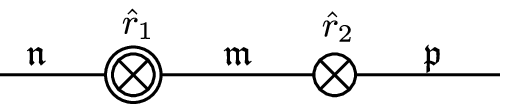}}}\,,
\ee
related to the weights by $\hat r_1=a_\groupn+p_a+p$, $\hat r_2=b_\groupm+c_1$.

\begin{figure}
\begin{centering}
\includegraphics[width=0.5\textwidth]{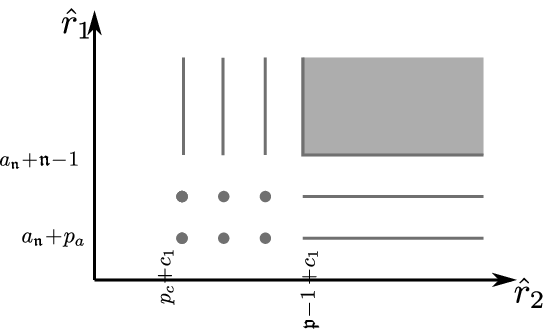}
\caption{\label{fig:r1r2}Possible places of unitarity for $\su(\groupn,\!|\groupm|\groupk)$ algebra.}
\end{centering}
\end{figure}

If the bosonic Dynkin labels are fixed, the values of $\hat r_1$ and $\hat r_2$ that are expected to give unitary representations are summarized in Fig.~\ref{fig:r1r2}.

The reason that we used "expected to give" is the following. As follows from  \cite{JakobUHW}, possible places of unitarity shown in Fig.~\ref{fig:r1r2} are exactly those that follow from the first approximation analysis - requirement that action of $\su(\groupn,\!|\groupm)$, $\su(\groupm|\groupk)$, $\su(\groupn,\groupk)$ subalgebras (separately each) on the highest weight is unitarizable. Subsequent more delicate analysis in \cite{JakobUHW} lead to the conclusion that some of the discrete places of Fig.~\ref{fig:r1r2} are in fact not unitary. This however contradicts the results of Dobrev and Petkova \cite{Dobrev:1985qv} for $\su(2,\!|\groupn|2)$ from which it follows that Fig.~\ref{fig:r1r2} describes exactly all UHW. For the moment the reason for this contradiction is not clear.\com{In particular, statement in \cite{JakobUHW} lead to the conclusion that the 1/2 BPS representation in $\psu(2,|4|2)$ (described by the labels $\langle 0001000\rangle$) is not unitary. Before this contradiction is clarified, one cannot put conjecture about bijection into a theorem. We expect though that the bijection conjecture is correct since the string hypothesis analysis so naturally leads to Fig.~\ref{fig:r1r2}.}

\paragraph{UHW for $\su(\groupn,\groupm|\groupk,\groupr)$} According to \cite{JakobUHW}, there is no UHW for the standard choice of Borel decomposition. The proof is based on the requirement of simultaneous unitarity of $\su(\groupn,\groupm)$, $\su(\groupk,\groupr)$, $\su(\groupm|\groupk)$, $\ldots$ submodules generated from the highest weight vector.

\section{\label{sec:conclusions}Conclusions and outlook}
In this article we derived a set of functional equations (\ref{onesimpleformofmaineequation}) which summarizes\footnote{On the level of equations, not solutions.} most of our knowledge about thermodynamic limit of unitary integrable rational spin chains of $gl(n|m)$ type. We propose to think about these equations as a suitable unification point for an analytical analysis of such spin chains, this is also a suitable place to begin derivation of the thermodynamic Bethe Ansatz equations, or Y-system supplemented with analytical properties of Y-functions\footnote{Necessity to solve the AdS/CFT spectral problem at finite volume increased an interest in the understanding of the  origins of the Y-system. Here we propose an analytic look on the problem, in the literature there are also different algebraic approaches  \cite{Derkachov:2008aq,Kazakov:2010iu,Bazhanov:2010jq}.}.
Below we list some  directions in which our work can be useful.\\

The set of equations (\ref{onesimpleformofmaineequation}), or equivalently (\ref{mainequation}), is the most symmetric way to formulate the thermodynamic limit of the Bethe Ansatz. For instance, either we study XXX spin chain or Gross-Neveu model, we get the same equations (\ref{onesimpleformofmaineequation}) with only difference in the source term. Therefore (\ref{onesimpleformofmaineequation}) make manifest  relations between different models. And indeed, chiral Gross-Neveu model can be viewed as a special limit of inhomogeneous  XXX\ spin chain. It would be interesting to collect all known continuous limits in the literature and classify them using the pattern of (\ref{onesimpleformofmaineequation}).

An unfortunate drawback is that string hypothesis, at least how we formulate it here, is not valid for the alternating supersymmetric spin chains \cite{Essler:2005ag}. We hope however that at least in some reduced form our analysis would be useful also for this case.\\

\begin{figure}
\begin{centering}
\includegraphics[width=0.4\textwidth]{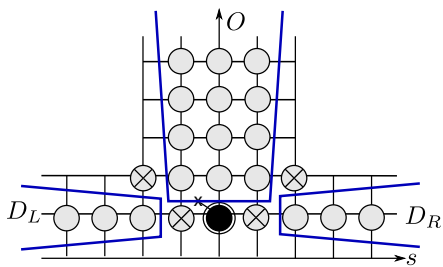}
\caption{\label{fig:gkphook} Organization of excitations around GKP string described \cite{Basso:2010in} into a T-hook domain of \cite{Gromov:2009tv}. Domain $O$ corresponds to isotopic $SO(6)$ excitations, domains $D_L$, $D_R$ - to the excitations in the AdS space, fermionic nodes - to fermions. Black node is the place were particle/hole transformation is made.}
\end{centering}
\end{figure}

Recent studies of scattering amplitudes \cite{Gaiotto:2010fk} showed  that it is important to understand structure of excitations around the GKP string.  Identification of these excitations was done  by Basso \cite{Basso:2010in}, and as a  potential application of (\ref{onesimpleformofmaineequation}) we show in Fig.~\ref{fig:gkphook} how the excitations found in \cite{Basso:2010in} are mapped to the nodes of the AdS/CFT T-hook pattern. Basically, a particle/hole transformation is made on the node $\{1,0\}$.  Immediately from Fig.~\ref{fig:gkphook} we see that $SO(6)$ isotopic excitations are indeed coming from the Bethe Ansatz equations for $A_3$ Dynkin diagram. Each type of Bethe nodes has one source term - the middle node of $A_3$ has $\{1,0\}$ particles (after particle/hole transformation $R_{1,0}$ becomes a resolvent for particles) as a source and left a right nodes have fermions as a source. For the excitations in the right and left strips of T-hook - it is clear that interaction between noncompact excitations goes in the same way as it is for $XXX_{1/2}$ spin chain, however interaction with fermions goes through holes, not through particles. Therefore, necessity to express holes in terms of particles makes Bethe equations for nodes in these strips more complicated.
\\

Exponentiation property (\ref{YK}) gives us a shortcut way for formulation of the Y-system. This observation leads to interesting questions for the AdS/CFT case. The AdS/CFT system at finite coupling implies introduction of additional terms in (\ref{onesimpleformofmaineequation}) for the nodes $\{1,0\}$, $\{1,\pm 1\}$, and $\{2,\pm 2\}$. Since in the AdS/CFT case the choice of monodromy used to define $K*\rho$ in (\ref{mainequation}) is not unique, one can get more from the functional equations (\ref{onesimpleformofmaineequation}) than a standard form of the Y-system (\ref{Ystandard}). Presumably, study of other monodromies should bring us to results equivalent to \cite{Cavaglia:2010nm}. From the other side, there is a belief \cite{Gromov:2009tv} that the standard form of the Y-system together with simple physical requirements is enough to find the finite volume spectrum. If that is true, results of \cite{Cavaglia:2010nm} should be derivable from physical requirements and therefore finite coupling modification of (\ref{onesimpleformofmaineequation}) should be derivable as well. This means in particular a new way to understand the origin of the dressing factor in the Beisert-Staudacher Bethe Ansatz equations.\\

A different puzzling feature of the patterns in Fig.~\ref{fig:patterns}, derived in this paper from an analytical study of the Bethe Ansatz equations, is that these patterns "know" somehow about unitary representations of the symmetry algebra. We conjectured that all unitary representations can be represented as inscription of certain generalized Young diagram inside a T-hook domain. This inscription rule applied for the rectangular representations is consistent   with  bilinear identities on the characters \cite{Gromov:2010vb} quantum version of which are the Hirota equations on the transfer matrices (\ref{Hirotaintro}).

It is important to notice that there is no well clarified explanation why the string hypothesis gives us knowledge about unitary representations. It is tempting to think that string configurations correspond to bound states in a given rectangular representation, but explicit realization of this idea is not obtained so far.

Notice also that we can define physically two different objects: $T$-functions which are defined through the $Y$-functions by the relation
\be
        Y_{a,s}=\frac{T_{a,s+1}T_{s,s-1}}{T_{a+1,s}T_{a-1,s}}\,,
\ee
and the transfer matrices (which are also denoted as $T_{a,s}$). These physically very different objects satisfy however exactly the same equation (\ref{Hirotaintro}) and the labels $\{a,s\}$ belong to exactly the same domain. In few cases we can make explicit computations to show that these objects coincide  though without real physical understanding why this happens. It would be very important to clarify the reasons for this miracle bijection.

\bigskip
\bigskip
\bigskip
\leftline{\bf \large Acknowledgments}
\

The author is grateful  to Benjamin Basso, Constantin Candu, Murat Gunaydin, Vladimir Kazakov, Ivan Kostov, Andrii Kozak, Carlo Meneghelli, Radu Roiban, Didina Serban,  Matthias Staudacher, Zengo Tsuboi, and Pedro Vieira  for valuable discussions. The author is especially grateful to Andrii Kozak for collaboration      in the initial stages of the project. This work was supported by the US
Department of Energy under contracts DE-FG02-201390ER40577.

\appendix

\section{\label{applicability}\label{sec:applicability}Applicability of string hypothesis}
String hypothesis was always considered as a plausible approximation as it passes the check of completeness - explicit counting of states can be shown to give the dimension of the Hilbert space of the spin chain model \cite{Bethe:1931hc,BetheCompleteness,Kirillov,KKR}. However, in literature there are known examples when string hypothesis is not satisfied, even qualitatively \cite{Isler:1993fc,Ilakovac:1999pe}. The most severe deviation appears when position $u$ of Bethe roots scales as $u\sim L$ with the volume $L$ of the spin chain. This case was in great detail analyzed for XXX spin chain in \cite{Bargheer:2008kj}.

Though the string hypothesis is violated, the numerical studies \cite{Antipov2006}  show that most of the Bethe Ansatz solutions obey the string configuration pattern. In this appendix we would make a study that supports this  point of view. An analytical treatment of $2$-string solution for XXX spin chain is made here, and this gives us a basic idea for what regimes string hypothesis can be used.

We expect that $2$-string solution would be a good approximation near origin while far from origin deviations from exact string configuration should be severe. The best way to track how string approximation depends on the distance to the origin is to make moving the center of the string. And this is indeed possible by using the twist in the Bethe Ansatz equations.

Let us add twist $\phi$ to the Bethe Ansatz equations (\ref{twistedBetheXXX}) and take logarithm. This leads to
\be\label{logBethe}
         \frac{L}{2\pi i} \log\frac{u_k+i/2}{u_k-i/2}-\frac 1{2\pi i}\sum_{j=1,j\neq k}^{M}\log\frac{u_k-u_j+i}{u_k-u_j-i}=n_k-\frac{\phi}{2\pi},
\ee
where $n_k$ is some integer called mode number.

By varying value of $\phi$ we can continuously change location of the Bethe roots and in this way change mode numbers $n_k$ of a solution. Large mode numbers correspond to the Bethe roots near origin while small mode numbers correspond to large Bethe roots. In particular, the point $n_k-\frac{\phi}{2\pi}=0$ corresponds to the Bethe root at infinity.

We focus now on the case $M=2$ and introduce the following notations:
\be
u_1=u_0+i/2+i\d,\ \ \ u_2=u_0-i/2-i\d.
\ee

String configuration can be never exact except at the origin since otherwise logarithmic singularity in (\ref{logBethe}) appears.
 Therefore $2$-strings separate into two classes: $2^+$-strings (with $\d>0$) and $2^-$-strings (with $\d<0$). Interchange between these two classes may happen only at origin.

Since two Bethe roots are complex conjugated to each other, they should have the same mode numbers. Definition of the mode number is unambiguous in the case of the $2^+$-strings within standard choice of the logarithm branch cut. For $2^-$-strings we define interaction between Bethe roots using  the formal rule $\log a\equiv {\rm Re}(\log a)$ if $a<0$.  This makes mode numbers half-integer.

Consider first the behavior of the $2$-string when $|\d|$ is small. Assuming that $|\d|$ is much smaller than $u$ we get
\be\label{smalldelta}
|\d|\sim \(\frac{u_0^2}{1+u_0^2}\)^{L/2}\simeq
\left\{
\begin{array}{cc}
u_0^L,&u_0\ll 1 \\
e^{-\frac{L}{2u_0^2}},& u_0\gg 1 \\
\end{array}\right. .
\ee
So deviation from exact string is  exponentially small in $L$ up to $u_0$ of scale $\sqrt{L}$.

If $u_0\ll 1$,   equations (\ref{logBethe}) are approximated by
\be
u_0=\frac {2\pi}{L}\(\frac L4\sign(u_0)-n_k+\frac \phi{2\pi}\).
\ee
We see that the maximal possible mode number is $\frac 12[\frac L2]$, where $[x]$ means integer part of $x$. For even values of $L$ there is an exact string at origin which corresponds to a highest vector state of the XXX spin chain. For even values of $L$ $2^+$- and $2^-$-strings do not interchange while for odd values of $L$ they do.

\begin{wrapfigure}{r}{0.4\textwidth}
\vspace{0pt}
\begin{center}
\parbox{0.35\textwidth}{
\includegraphics[width=0.35\textwidth]{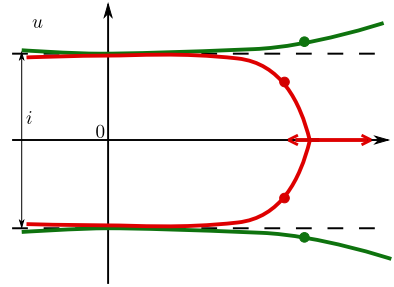}
}
\end{center}
\caption{\label{fig:twistdependence}\footnotesize  Trajectories traced by $2^+$ (green) and $2^-$ (red) strings when the twist is continuously changed.}
\vspace{-15pt}
\end{wrapfigure}
Next, consider the behavior of the $2$-string for large $u_0$. When $n_k-\frac \phi{2\pi}$ approaches zero $u_k$ should become not only large, but also far from other Bethe roots. This means that $\d$ of $2^+$-string should become large. Indeed, assuming that $u_0\gg 1$ and $\d\gg1$ we get
\be\label{largedelta}
u_0\sim\frac{L-1}{2\pi n_k-\phi},\ \ \delta\sim\frac{\sqrt{L-1}}{4\pi n_k-2\phi}.
\ee

$2^+$-string significantly  deviates from an exact $2$-string at $(2\pi n_k-\phi)\sim\ \sqrt{L-1}$, {\it i.e.} at scales $u\sim\sqrt{L}$ or larger. This is consistent with the analysis at small values of $|\d|$.

For $2^-$-string $|\d|$ cannot become large since we cannot pass through $\d=0$ point and negative values of $\d$ are bounded by $-1/2$. The only possibility for the two Bethe roots of $2^-$-string to become far from one another is to become two real Bethe roots\footnote{This is precisely the situation observed in \cite{Isler:1993fc}. We see that introduction of twist is an effective way to predict this situation. }. For this they should first collide at the real axis and then separate. After separation we no more need prescription for $\log a$ with $a<0$, this effects into separating of two half-integer mode numbers into two subsequent integers. In this way two Bethe roots will approach infinity with the interval $2\pi$ of $\phi$.

In summary, we expect the qualitative dependence of $2$-strings on the twist as shown in Fig.~\ref{fig:twistdependence}. We support our  study with  numerical calculations done in {\it Mathematica}, the code is the following\footnote{The twist is made slightly complex to avoid the collision point }:

\medskip
\medskip
\medskip
\begin{minipage}{0.8\textwidth}
\footnotesize{
\begin{verbatim}
BAE[f_]:= Table[((u[k]+I/2)/(u[k]-I/2))^L Exp[2 Pi I f] ==
         -Product[(u[k]-u[i]+I)/(u[k]-u[i]-I), {i,M}], {k,M}];
(* choose one of the initial conditions *)
(*1*) L=11;M=2;init={u[1]->0.457-0.501 I,u[2]->0.457+0.501 I};
(*2*) L=11;M=2;init={u[1]->0.88 - 0.49 I,u[2]-> 0.88+ 0.49 I};
(*3*) L=8; M=3;init={u[1]->-0.49- 0.99 I,u[2]-> -0.50,
                     u[3]->-0.49+ 0.99 I};
step = 1/100; cycle = step Range[1/step]; ncycles = 2;
path = (Join @@ Table[cycle+k, {k, 0, ncycles - 1, 1}])+I/100;
str[path[[1]]] =
  FindRoot[BAE[path[[1]]],Table[{u[k], u[k]/.init}, {k, M}]];

Do[str[path[[r]]] = FindRoot[BAE[path[[r]]], Table[ {u[k],
        u[k]/.str[path[[r-1]] ]},{k, M}],
   WorkingPrecision -> 15]; NotebookDelete[temp];
 temp = PrintTemporary[path[[r]] // N], {r, 2, Length[path]}];

Manipulate[
{ListPlot[{Re[#],Im[#]}&/@Table[u[k]/.str[path[[r]]], {k, M}],
   PlotRange -> {{-4, 4}, {-1.3, 1.3}},
   PlotStyle -> {Red, PointSize[.04]}],
   SetAccuracy[path[[r]], 4]}, {r, 1, Length[path], 1}]
\end{verbatim}
}
\end{minipage}
\medskip
\medskip

The main conclusion of the analysis is that string approximation works well for $u_0\lesssim\sqrt{L}$ and is inappropriate otherwise. At large values of $u$ mode numbers can be approximated by $n_k\sim L/(2\pi u_k)$. Therefore among all possible complex solutions only $1/\sqrt{L}$ fraction of them significantly deviates from the exact $2$-string configuration. For ferromagnetic case however this $1/\sqrt{L}$ fraction is important since it corresponds to the lowest-energy excitations. For antiferromagnetic case the solutions which deviate significantly from the exact string are highly-energetic and therefore inessential. We conclude that for antiferromagnetic study in the large volume string hypothesis is a correct approximation

We assume that the effects observed on the example of $2$-string solution are similar in a general situation, in particular approximations (\ref{smalldelta}) and (\ref{largedelta}) seem not to change significantly. This leads to the assumption formulated in section \ref{sec:status} as the weak string hypothesis.

\section{\label{sec:numerical}A numerical example for string objects}
This appendix aims to give a numerical evidence that more complicated objects like strings on the nested levels and triangle/trapezium-type configurations do exist.

First, we consider $\su(3)$ spin chain and we will present a string configuration on the nested level. The set of the Bethe equations is given by

\medskip
\medskip
\begin{minipage}{0.85\textwidth}
\footnotesize{
\begin{verbatim}
BAE := Table[((u[k]+I/2)/(u[k]-I/2))^L Product[(u[k]-v[i]+I/2)/
             (u[k]-v[i]-I/2),{i,M2}] == -Product[(u[k]-u[i]+I)/
             (u[k]-u[i]-I), {i,M}], {k, M}];
NestedBAE[f_] := Table[Product[ (v[k]-u[i]+I/2)/(v[k]-u[i]-I/2),
            {i, M}]Exp[2 \[Pi] I f] == -Product[(v[k]-v[i]+ I)/
            (v[k]-v[i]-I), {i,M2}], {k,M2}];
allBAE[f_] := BAE~Join~NestedBAE[f];
\end{verbatim}
}
\end{minipage}
\medskip
\medskip

To find an approximative solution we introduce the twist on the nested level and choose it to be small and negative. This allows us to choose such Bethe roots on the nested level which are large comparatively to the Bethe roots of the momentum-carrying node. For the momentum-carrying node we choose a configuration with all Bethe roots real. For the nested level we choose a string configuration which can be approximated using zeroes of Hermite polynomial \cite{Bargheer:2008kj}:

\begin{minipage}{0.85\textwidth}
\footnotesize{
\begin{verbatim}
\end{verbatim}
}
\end{minipage}

\begin{minipage}{0.85\textwidth}
\footnotesize{
\begin{verbatim}
L = 108; M = 26; M2 = 5; initf = -1/10; prec = 10;
Do[n[k]=If[k > M/2,-Round[L/4]-M+2 k,Round[L/4]-2 k],{k,M}];
BAElog := Table[L Log[(u[k]+I/2)/(u[k]-I/2)] ==
    Sum[If[i==k,0,Log[(u[k]-u[i]+I)/(u[k]-u[i]-I)]], {i,
       M}] + 2\[Pi] I n[k], {k, M}];
initu = FindRoot[BAElog, Table[{u[k], L/(2 \[Pi] n[k])}, {k, M}],
   WorkingPrecision -> prec];
initv = Table[v[k]->(M+I Sqrt[2M]Root[HermiteH[M2,#]&,k]//N)/
                 (2 \[Pi] (-initf)), {k, M2}];
\end{verbatim}
}
\end{minipage}

\medskip
\medskip

Then, iterating in twist one can bring  the solution to the region where it approaches the exact string configuration.

\medskip
\medskip
\begin{minipage}{0.85\textwidth}
\footnotesize{
\begin{verbatim}
step = 1/40; path = Table[k, {k, initf, -2, -step}];
sol[path[[1]]] =  FindRoot[allBAE[initf],
     Table[{u[k], u[k] /. initu}, {k, M}]~Join~
     Table[{v[k], v[k] /. initv}, {k, M2}],
     WorkingPrecision -> prec] // Chop;
Do[sol[path[[r]]] = Chop[#, 10^(-prec+4)] &@
    FindRoot[allBAE[path[[r]]],
     Table[{u[k], u[k] /. sol[path[[r - 1]]]}, {k, M}]~Join~
     Table[{v[k], v[k] /. sol[path[[r - 1]]]}, {k, M2}],
     WorkingPrecision -> prec];
     NotebookDelete[temp];temp = PrintTemporary[path[[r]] // N],
     {r, 2, Length[path]}]
\end{verbatim}
}
\end{minipage}

\medskip
\medskip

\begin{minipage}{0.85\textwidth}
\footnotesize{
\begin{verbatim}
rep = Rule[a_, b_] :> b;
Manipulate[{ListPlot[
         {{Re[#],Im[#]}&/@sol[path[[r]]][[1;;-M2- 1]]/.rep,
          {Re[#],Im[#]}&/@ sol[path[[r]]][[-M2;;-1]]/.rep},
           PlotRange -> {{-10, 10}, {-2.3, 2.3}},
    PlotStyle -> {{Red, PointSize[.04]}, {Blue, PointSize[.03]}}],
          SetAccuracy[path[[r]], 4]},
{r, 1, Length[path], 1}]
\end{verbatim}
}
\end{minipage}
\medskip
\medskip

The positions of $v$-roots are given by:

\medskip
\medskip
\begin{minipage}{0.85\textwidth}
\footnotesize{
\begin{verbatim}
In:=   v /@ Range[M2] /. sol[path[[-1]]] // N
Out:=  {1.79704 - 1.95446 I, 1.81005- 0.959531 I,
  1.81005+ 0.0404624 I,  1.81004+ 1.04047 I, .79661+ 2.03238 I}
\end{verbatim}
}
\end{minipage}
\medskip
\medskip

\noindent which is a fairly good string configuration.

A trapezium or triangle configuration can be simply obtained using a duality transformation. Let us take the above $\su(3)$ example and treat it as a solution of the $\su(3|1)$ (\xOOX) Bethe Ansatz equations with no Bethe roots at the fermionic node. Now, let us perform the duality transformation on the fermionic node which will give us Bethe equations for \xOXX. The Bethe roots of the dualized fermionic node are given by:

\medskip
\medskip
\begin{minipage}{0.89\textwidth}
\footnotesize{
\begin{verbatim}
In:=     Q2[u_] = Product[(u - v[k]), {k, M2}] /. sol[path[[-1]]];
         u /. NSolve[Q2[u + I/2] - Q2[u - I/2], u] // N
Out:=    {1.7993+ 1.53399 I, 1.79964- 1.45547 I, 1.81004+ 0.540469 I,
             1.81005- 0.459534 I}
\end{verbatim}
}
\end{minipage}
\medskip
\medskip

We see that $v$-roots and the Bethe roots of the dualized fermionic node form a $(5,4)$-trapezium configuration. The same reasoning can be continued further, and for this particular example  trapezia configurations up to a $(5,1)$-trapezium ($5$-triangle) for the $\su(3|4)$ algebra can be obtained.

\medskip
\medskip
\begin{minipage}{0.85\textwidth}
\footnotesize{
\begin{verbatim}

\end{verbatim}
}
\end{minipage}

\section{\label{sec:sourcewing}Source term in the wing.}
Construction of the main text allows putting the source term in the arbitrary node of the vertical strip in T-hook. It would be natural to expect the situation when the source term is  inside right or left wings. This indeed can be done, however one should consider representations with more than one nonzero Dynkin labels.

\begin{figure}[t]
\begin{centering}
     {\includegraphics[width=0.8\textwidth]{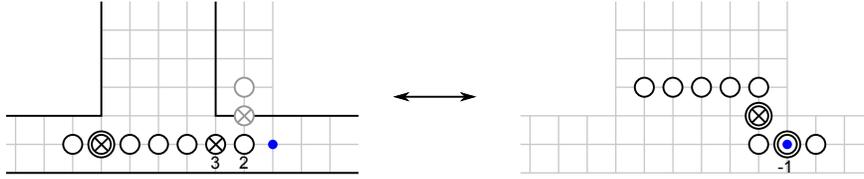}}
     \caption{\label{fig:sourcewing}Two different Kac-Dynkin labelings serving the same set of functional equations. On the right the situation is the same as in main text. On the left we have two source terms. By demanding absence of excitations at two Dynkin nodes depicted by gray on the left, we get $\su(2,2|4)$ T-hook with a source term in the right wing.}
\end{centering}
\end{figure}

To be precise, let us consider $\su(2,2|4)$ case and try to find the situation which corresponds to the source term marked by the blue dot in Fig.~\ref{fig:sourcewing}. Expected is  to get this case from the representation with Dynkin labels $\langle 0000032\rangle$. To be able apply the general pattern, we should add two more Dynkin nodes (with zero number of excitations on them) and perform duality transformation to get the Kac-Dynkin diagram with only one nonzero Dynkin label\footnote{notations here are the same as in Fig.~\ref{fig:duality}.}:
\be\label{casehookedge}
\parbox{0.5\textwidth}{\includegraphics[width=0.48\textwidth]{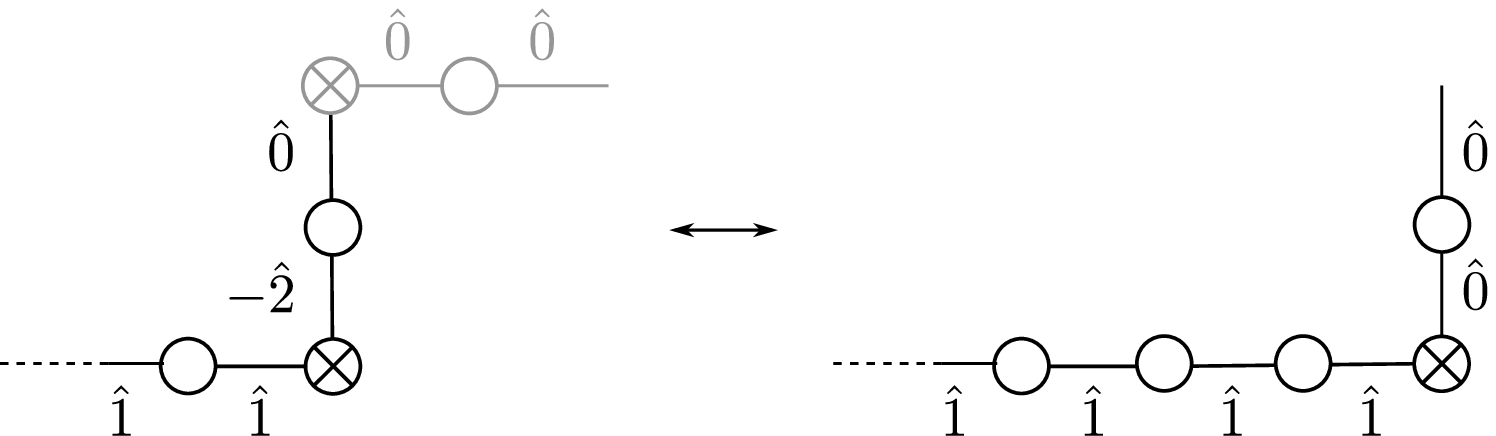}}
\ee

It is possible (but not compulsory) to perform duality transformations further and get the distinguished Dynkin diagram, the functional equations for which are encoded on the right part of Fig.~\ref{fig:sourcewing}. These functional equations obviously contain the source term at the desired position. Then, by undoing the duality transformations and suppressing auxiliary Dynkin nodes one obtains the $\su(2,2|4)$ T-hook.

The case when the source term is on the boundary of T-hook is also possible. We can treat it similarly, by introducing the auxiliary Bethe equations. The only difference is that when we rewrite equations in a form that do not contain auxiliary resolvents, the source term would not enter (\ref{onesimpleformofmaineequation}). Instead, it will appear in the corner equation (\ref{cornerequation}). We leave to the reader derivation of how the corner equation should be modified in this particular case.

\begin{figure}[h]
\begin{centering}
\includegraphics[width=0.4\textwidth]{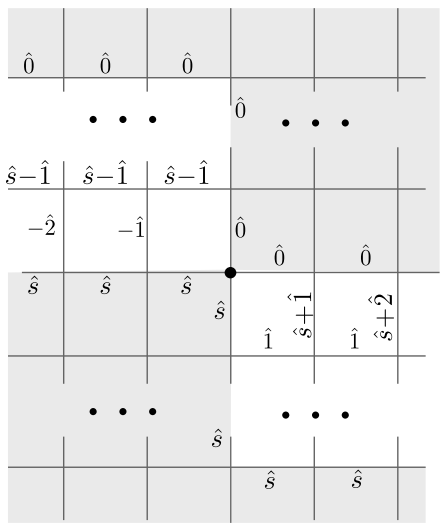}
\caption{\label{fig:dynkinlattice}Lattice of all possible weights for different choices of the Kac-Dynkin diagrams. In the grey regions all the  weights are equal to $\hat 0$ or $\hat s$. We draw the Kac-Dynkin diagram as a ladder from bottom left to top right turning at the fermionic nodes. So, the diagram has nonzero Dynkin labels only when it passes through white regions or the center. In particular, maximally only two bosonic nodes can have nonzero Dynkin labels, at the places where the diagram crosses the borders of the gray regions. }
\end{centering}
\end{figure}

One may ask, in what situation it is possible to obtain, using duality transformations and auxiliary Dynkin nodes, the grading in which only one Dynkin label is not zero. To answer this question, let us assume that we've got already this grading and that nonzero label is  equal to $\hat s$,  positive for simplicity. Then, by applying the rule of Fig.~\ref{fig:duality}, we get a lattice with numbers assigned to each edge. These numbers are possible weights of $\hat E_{ii}$ generators\footnote{vertical lines correspond to the generators with $|i|=1$ and horizontal - with $|i|=0$.}. We show this lattice in Fig.~\ref{fig:dynkinlattice}.

The Dynkin diagrams which can be drawn on this lattice constitute the full set of the diagrams reducible to the one with only one nonzero Dynkin label. In particular, it is easy to see how to embed the case (\ref{casehookedge}) in  Fig.~\ref{fig:dynkinlattice} with $\hat s=1$.

\newpage
\bibliography{phdbib}        
\bibliographystyle{utphys}

\end{document}